\documentclass[iop]{emulateapj}

\renewcommand{\deg}{^{\circ}}

\newcommand{\ecr}{\varepsilon_{cr}}

\usepackage[colorlinks,urlcolor=blue,citecolor=blue,linkcolor=blue]{hyperref}

\shorttitle{Simulations of ion acceleration at shocks: DSA efficiency}
\shortauthors{Caprioli \& Spitkovsky}

\begin{document}

\title{Simulations of ion acceleration at non-relativistic shocks:\\
i) Acceleration efficiency}

\author{D. Caprioli and A. Spitkovsky}
\affil{Department of Astrophysical Sciences, Princeton University, 
    4 Ivy Ln., Princeton NJ 08544}
\email{caprioli@astro.princeton.edu}

\begin{abstract}
We use 2D and 3D hybrid (kinetic ions -- fluid electrons) simulations to investigate particle acceleration and magnetic field amplification at non-relativistic astrophysical shocks.
We show that diffusive shock acceleration operates for quasi-parallel configurations (i.e., when the background magnetic field is almost aligned with the shock normal) and, for large sonic and Alfv\'enic Mach numbers, produces universal power-law spectra $\propto p^{-4}$, where $p$ is the particle momentum. 
The maximum energy of accelerated ions increases with time, and it is only limited by finite box size and  run time.
Acceleration is mainly efficient for parallel and quasi-parallel strong shocks, where 10--20\% of the bulk kinetic energy can be converted to energetic particles, and becomes ineffective for quasi-perpendicular shocks.
Also, the generation of magnetic turbulence correlates with efficient ion acceleration, and vanishes for quasi-perpendicular configurations.
At very oblique shocks, ions can be accelerated via shock drift acceleration, but they only gain a factor of a few in momentum, and their maximum energy does not increase with time.
These findings are consistent with the degree of polarization and the morphology of the radio and X-ray synchrotron emission observed, for instance, in the remnant of SN 1006.
We also discuss the transition from thermal to non-thermal particles in the ion spectrum (supra-thermal region), and we identify two dynamical signatures peculiar of efficient particle acceleration, namely the formation of an upstream precursor and the alteration of standard shock jump conditions.
\end{abstract}


\section{Introduction}\label{sec:intro}
The problem of finding the sources of the energetic extraterrestrial nuclei that we call cosmic rays (CRs) is a long-lasting one. 
\cite{baade-zwicky34} proposed supernova remnants (SNRs) to be responsible for such non-thermal particles, requiring about 10--30\% of the kinetic energy of the SN ejecta to be channelled into accelerated particles.
This energetic argument, coupled with the acceleration mechanism put forward by \cite{Fermi49,Fermi54}, was at the basis of the theory of \emph{diffusive shock acceleration} (DSA) developed in the late '70s by several authors \citep{krymskii77,axford+78,bell78a,bell78b,blandford-ostriker78}.
The most important feature of the DSA mechanism is that the spectrum of the accelerated particles is predicted to be a universal power-law, with a spectral index that depends only on the shock compression ratio.
Since for strong (i.e., large Mach number) shocks the compression ratio tends to the asymptotic value of $r=4$, particles are expected to be accelerated with a spectrum $f(p)\propto p^{-3r/(r-1)}\propto p^{-4}$, with $p$ the modulus of the particle momentum \citep[see, e.g.,][]{blandford-ostriker78}.  

This universality is crucial to account for the spectrum of Galactic CRs observed at Earth, which extends as a power-law $\propto E^{-2.7}$ from a few GeV up to a few times $10^6$ GeV for protons, and up to an additional factor $Z$ for heavier nuclei with charge $Z$.
The discrepancy between the spectrum predicted by DSA ($\propto E^{-2}$ in energy for relativistic particles) and the measured one can be explained by accounting for the CR transport in the Galaxy (the residence time is inferred to be a decreasing function of the energy) and for the differential escape from the sources \citep[e.g.,][]{pz05,nuclei}.
Moreover, radio observations of SNRs suggest that the energy distribution of accelerated electrons (in the 1--10 GeV range) is consistent with the DSA prediction \citep[e.g.,][]{Trushkin98}. 
Also, $\gamma$-rays from SNRs are often interpreted to be of hadronic origin \citep[see, e.g.,][]{gamma,pionbump}, suggesting that nuclei acceleration can be efficient at SNR forward shocks \citep[about 10\% in Tycho, see][]{tycho}.

Understanding particle acceleration at non-relativistic shocks, and the conditions that may favor it, is not a problem limited to SNRs, though.
Most astrophysical shocks are \emph{collisionless}, i.e., energy and momentum conversion is not mediated by interparticle collisions, the mean free path for Coulomb scattering being much larger the size of the system, but rather by collective electromagnetic processes.
There are many examples of collisionless shocks besides SNR ones, in a very wide range of scales: in the Solar System (e.g., the Earth's bow shock, the interplanetary shocks triggered by coronal mass ejections, and the solar-wind-related shocks), at the radio lobes of the jets of active galaxies, and also in clusters of galaxies.
These shocks span a wide range of sonic Mach numbers, magnetization, and relative inclination between the shock velocity and the unperturbed magnetic field, but they are typically associated with some form of non-thermal emission, which attests to the presence of accelerated particles.

In many cases, astrophysical shocks are also associated with magnetic fields much larger than the ambient ones, up to levels that cannot be accounted for by simple compression.
For instance, X-ray observations of young SNRs lead to inferred magnetic fields as large as a few hundreds $\mu$G, a factor of 50--100 larger than in the interstellar medium \citep[see, e.g.,][and references therein]{V+05,P+06}. 
Such amplified fields are likely produced by accelerated ions via different plasma instabilities \citep[see, e.g.,][for a review]{Bykov+13}.

CR spectra, acceleration efficiency and magnetic field amplification have been calculated via a two-fluid approach \citep[][for a review]{drury83}, via Monte Carlo simulations \citep[e.g.,][]{EMP90,jones-ellison91,EBJ95,EBJ96,no04,veb06}, as well as by solving the CR transport equation either numerically \citep[e.g.,][]{bv97,bv04,kang-jones97,KJG02,kang-jones06,za10}, or analytically  \citep[][and references therein]{malkov97,blasi02,ab06,boundary,efficiency}.
These methods return very consistent results \citep{comparison}, and can tackle the large scale dynamics of the shock, also including the CR back-reaction. 
Nevertheless, they need to be fed with microphysical prescriptions for particle scattering and injection, and for the excitation of the magnetic turbulence.

To overcome these limitations and to self-consistently deal with the highly non-linear interplay between particles and electromagnetic fields, numerical kinetic simulations are needed.
Such simulations of non-relativistic collisionless shocks have been carried out either in the full particle-in-cell (PIC) approach \citep[see, e.g,][]{ah07,ah10,rs11,Niemiec+12}, or in the hybrid (kinetic ions--fluid electrons) approach \citep[see, e.g.,][and references therein]{winske85,quest88,Giacalone+93,be95,wo96,Giacalone+97,ge00,Giacalone04,Lipatov02,gs12,gg13}.
A novel approach to DSA exploiting Vlasov--Fokker--Plank simulations has also been recently proposed by \cite{bell+13}; 
such a method does not include a self-consistent description of the injection physics, but it represents a powerful tool for studying accelerated particles on large scales.

In this paper we want to test, by means of kinetic hybrid simulations, the spectrum of ions accelerated at non-relativistic collisionless shocks via DSA, also investigating under which conditions ion acceleration and magnetic field amplification are effective.
Compared to PIC simulations, the hybrid approach does not resolve the small electron plasma scales, allowing us to simulate more macroscopic systems without losing any information about the shock dynamics, which is mostly driven by ions.
Very importantly, the self-consistent treatment of the interplay between accelerated particles and electromagnetic fields allows us  to study the correlation between ion acceleration and magnetic field amplification.

We have performed two- and three-dimensional (2D and 3D) simulations of non-relativistic shocks with different strengths (parametrized by how supersonic and super-Alfv\'enic the upstream flow is), and with different orientations of the initial magnetic field.
The presented simulations cover an unprecedentedly large range of parameters that includes both weak and strong shocks, and parallel, oblique and quasi-perpendicular configurations. 
A preliminary investigation of such a parameter space has been performed, for instance, by \cite{Giacalone+97,gs12} (hereafter, GS12), but the present one considerably improves on the previous analyses in several respects. 

In particular, we improve on the work of GS12 by allowing for much larger computational boxes and much longer time evolution, for a much larger parameter space in shock strengths and inclinations.
We have also checked the results obtained in 2D simulations against 3D ones, also performing an extensive study of the dependence of the results on time resolution, which is crucial to account for the kinematics of high-energy particles, especially at high Mach numbers.
All of these ingredients are necessary to properly lead to the formation of \emph{a power-law tail, extending for orders of magnitude in the non-thermal regime with the slope predicted by the DSA theory}, which has previously not been reported in the literature. 
Moreover, for the first time we present evidence for \emph{CR-modified shocks}, i.e., shocks whose dynamics is relevantly affected by the pressure and energy density in the accelerated particles, which leads to very distinctive features like the formation of an upstream precursor and the modification of the standard jump conditions of  gaseous shocks.
The improved energy conservation and longer run times compared to GS12 work enables us to refine the measurements of the accelerating efficiency of shocks as a function of flow conditions. 

The paper is structured as follows.
In section \ref{sec:DSA} we outline the simulation technique and demonstrate that at strong parallel shocks the momentum spectrum of accelerated ions is perfectly consistent with the prediction of DSA theory ($\propto p^{-4}$), also showing that its extent increases with time.
In section \ref{sec:therm} we characterize the transition between thermal and non-thermal particles, in particular discussing the properties of the peculiar \emph{supra-thermal} region of the ion spectrum, while in section \ref{sec:eff} we illustrate the dependence of the CR acceleration efficiency on the shock Mach number and magnetic inclination. 
An important result is that, while at parallel shocks a fraction as large as 10--20\% of the shock ram pressure is converted into accelerated ions, very oblique shocks are inefficient in producing energetic particles and magnetic turbulence (section \ref{sec:MFA}).
The observational implications of these findings are also outlined, with special attention to the case of SN1006.
Sections \ref{sec:CRmod} and \ref{sec:SDA}  provide a description of the dynamics of the CR-modified shocks and of the role played by \emph{shock drift acceleration}, respectively.
The results of 3D simulations are shown in section \ref{sec:3d}.
We conclude in section \ref{sec:conclusions}.

\section{Diffusive shock acceleration}\label{sec:DSA}
\begin{figure*}\centering
\includegraphics[trim=2px 0px 2px 0px, clip=true, width=1\textwidth]{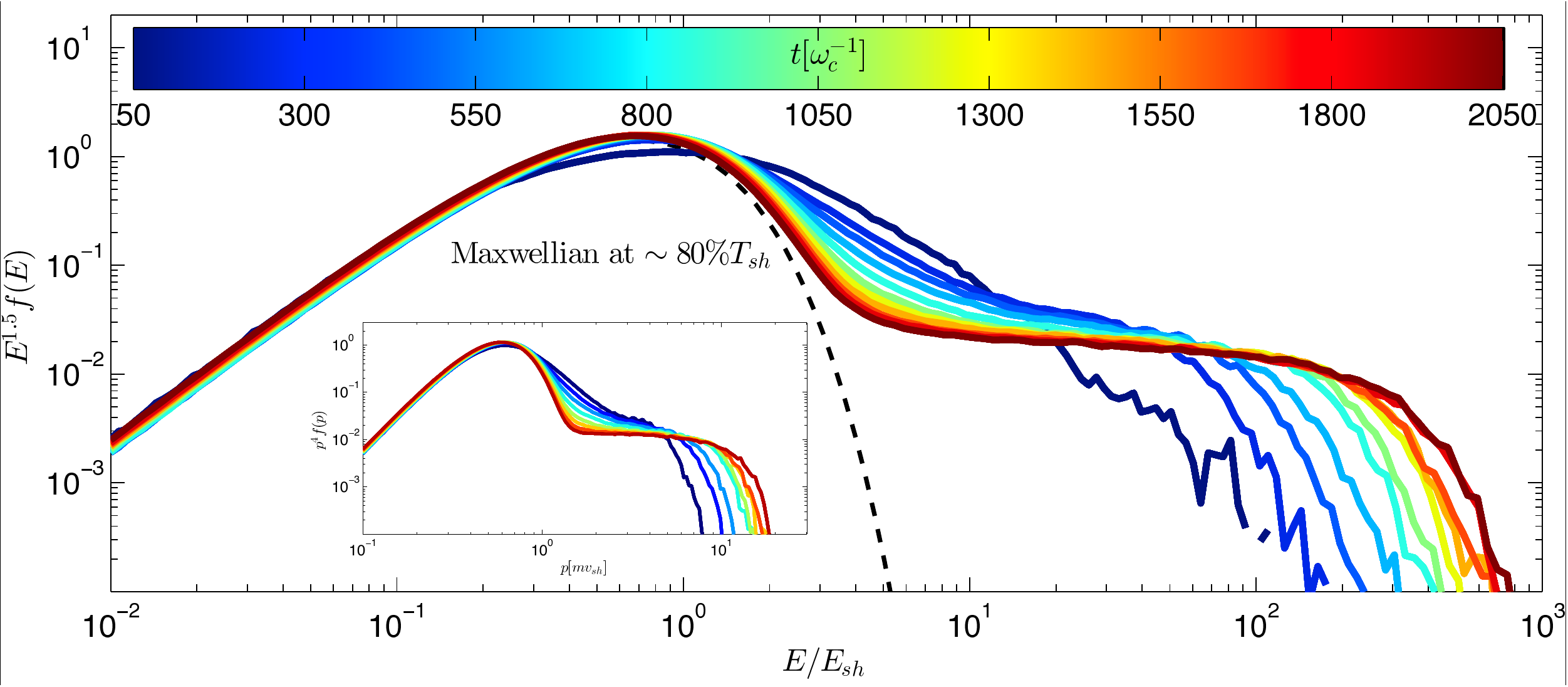}
\caption{\label{fig:evo}
Downstream ion energy spectrum at different times, as in the legend. 
It is possible to note the thermal distribution, well-fitted by a Maxwellian with temperature about 80\% the one expected for Mach number 20 shock that does not accelerate particles (dashed line), and a non-thermal power-law tail extending to larger energies at later times.
The spectrum is plotted multiplied by $E^{1.5}$, to emphasize the agreement with the energy scaling predicted by DSA, which reads $p^{-4}$ in momentum (see inset and eq.~\ref{eq:1}). 
}
\end{figure*}

The simulations in this paper are carried out with the \emph{dHybrid} code \citep{gargate+07}, a massively parallel, Newtonian (i.e., non-relativistic), hybrid code, in both its 2D and 3D configurations. 
Ions are treated kinetically and electrons as a neutralizing fluid, with a prescribed polytropic equation of state.
Lengths are measured in units of the ion skin depth $c/\omega_p$, where $c$ is the light speed and $\omega_p=\sqrt{4\pi n e^2/m}$ is the ion plasma frequency, with $m,e$ and $n$ the ion mass, charge and number density, respectively;
time is measured in inverse cyclotron times $\omega_c^{-1}=mc/eB_0$, with $B_0$ the strength of the initial magnetic field.
Finally, velocities are normalized to the Alfv\'en speed $v_A=B/\sqrt{4\pi m n}=c\omega_p/\omega_c$.
All simulations include the three spatial components of the particle momentum, and the three components of electric and magnetic fields. 

Let us consider a 2D shock with Alfv\'enic Mach number $M_A=v_{sh}/v_A=20$, where we define $v_{sh}$ to be the upstream fluid velocity in the downstream reference frame.
The initial magnetic field ${\bf B}_0=B_0{\bf x}$ is in the same direction of ${\bf v}_{sh}=-v_{sh}{\bf x}$ (parallel shock).
Ions are initialized with thermal velocity $v_{th}=v_A$, so that their temperature  reads $T_0 = \frac{1}{2}mv_{A}^2 / k_B$, with $k_B$ the Boltzmann constant. 
Electrons are initially in thermal equilibrium with ions, i.e., $T_e=T_i=T_0$, and their adiabatic index is chosen in such a way to reproduce the expected jump conditions at the shock, as in GS12.
The sound speed is $c_s=\sqrt{2\gamma k_BT_0/m}$ and hence the sonic Mach number reads $M_s=M_A\sqrt{\gamma}$, with $\gamma=5/3$ the ion adiabatic index.
This regime is typical of the cold interstellar medium: with $n=0.1$cm$^{-3}$, $B_0=3\mu$G and $T=10^4$K one has $v_A\approx c_s\approx 15$kms$^{-1}$.
Except when otherwise specified, throughout the paper we indicate the shock strength simply with $M=M_A\approx M_s$. 
The reader may refer to, e.g., \cite{Giacalone+97} for a survey of hybrid simulations of non-relativistic shocks with different sonic and Alfv\'enic Mach numbers.

The computational box measures $(L_x,L_y)=(10^5, 10^2) [c/\omega_p]^2$, with two cells per ion skin depth and 4 (macro)particles per cell; the time-step is chosen as $\Delta t=5\times 10^{-3}\omega_c^{-1}$.
The shock evolution is followed for many ion cyclotron times, until $t=2500\omega_c^{-1}$. 

The shock is produced by sending a supersonic flow against a reflecting wall (left side in figures);
the interaction between the initial stream and the reflected one produces a sharp discontinuity, which propagates to the right in the figures. 
As a consequence, in the simulation the downstream fluid is at rest, and the kinetic energy of the upstream flow is converted into thermal energy at the shock front.
It is worth mentioning that --- in the literature --- the shock Mach number is often quoted as measured in the shock reference frame, here indicated as $\tilde{M}$, which is related to $M$ through the implicit relation
\begin{equation}\label{eq:Mtilde}
\tilde{M}=M\left[ 1+\frac{1}{r(\tilde{M})}\right],\quad r=\frac{(\gamma+1)\tilde{M}^2}{(\gamma-1)\tilde{M}^2+2}
\end{equation}
namely, $\tilde{M}=5/4 M$ for a strong shock with $r=4$. 

The downstream ion energy distribution is shown in figure \ref{fig:evo}, as a function of time; we can identify three main spectral regimes.
At low energies, we have a thermal distribution, well-fitted with a Maxwellian (dashed line), the temperature of which --- at late times --- is $\sim 20\%$ lower than the temperature one would predict for a strong shock without CRs.
The main reason for this reduced heating of the downstream plasma is that about 20\% of the energy flux is channeled into non-thermal particles (also see section \ref{sec:CRmod} for more details on how CRs modify the global shock dynamics).

Always at late times, we see that the ion spectrum goes as $E^{-1.5}$ for $E\gtrsim 3 E_{sh}$, where we introduced 
\begin{equation}
E_{sh}=\frac{1}{2}mv_{sh}^2=\frac{1}{2}mM^2v_A^2.
\end{equation}
Such a power-law is in remarkable agreement with the DSA prediction for strong shocks \emph{for non-relativistic particles}, as we now discuss.
In a nutshell, the DSA mechanism relies on the fact that particles diffusing back and forth across the shock repeatedly gain energy because of first-order Fermi acceleration \citep{Fermi54}.
The spectrum of the accelerated particles does not depend on the details of the scattering, but only on the density jump between upstream and downstream, $r$.
For $M\gg1$ the shock compression ratio is $r\simeq 4$, and the spectrum of accelerated ions is predicted to be $\propto p^{-q}$ in momentum space, with $q=3r/(r-1)\simeq 4$.

The energy distribution $f(E)$ can be calculated as 
\begin{equation}\label{eq:1}
4\pi p^2 f(p)dp= f(E)dE \to f(E)=4\pi p^2 f(p)\frac{dp}{dE}.
\end{equation}
In the non-relativistic regime $E=p^2/2m$, so that $\frac{dp}{dE}\propto 1/p\propto E^{-1/2}$ and $f(E)\propto E^{-1.5}$;
in the relativistic limit, instead, $E\propto p$ and $f(E)\propto E^{-2}$.

In spite of the fact that simulations of non-relativistic collisionless shocks have been performed for many years (see section \ref{sec:intro}), this is the first time ---to our knowledge--- that the DSA prediction for strong shocks has been convincingly recovered in self-consistent simulations.
Previous simulations, while indeed showing evidence of supra-thermal ions and, occasionally, of power-law distributions, have never been run for long enough, and  in sufficiently large computational boxes, to unequivocally see ions accelerated through DSA over almost 3 decades in energy (figure \ref{fig:evo}).
Moreover, in this work we account for Mach numbers as large as 50, while most of the previous work has been done for shocks with $M\lesssim 10$; in such a regime the magnetosonic Mach number is $v_{sh}/\sqrt{c_s^2+v_A^2}\lesssim 7$, implying $r<4$ and, in turn, $q>4$. 

\begin{figure*}\centering
\includegraphics[trim=0px 0px 0px 0px, clip=true, width=.8\textwidth]{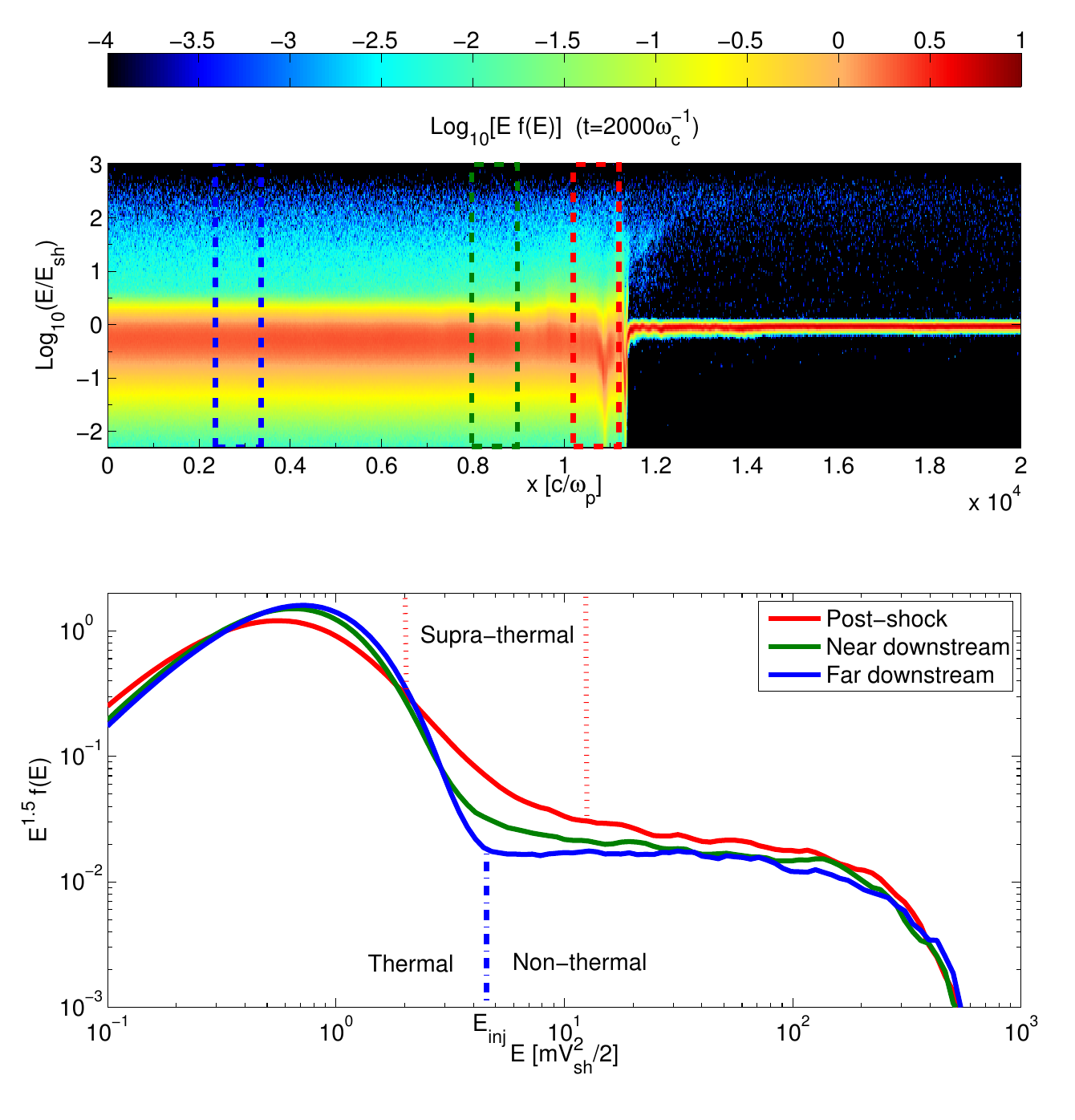}
\caption{\label{fig:therm}
\emph{Top panel:} ion energy spectrum (color code) as a function of $x$.
The transition from the cold beam to the broad distribution marks the shock position, at $\approx 1150c/\omega_p$. 
Note the population of high-energy ions diffusing ahead of the shock (for $E\gtrsim 10E_{sh}$).
\emph{Bottom panel:} ion energy distribution at three different downstream locations, corresponding to the dashed boxes in the top panel.
Immediately behind the shock (red curve) there is a ``bridge'' of supra-thermal particles smoothly connecting the thermal peak with the DSA power-law, while far downstream (blue curve) there is quite a sharp boundary between thermal and non-thermal ions at $E_{inj}\sim 3 - 4 E_{sh}$.
\emph{A color version is available in the online journal}.
}
\end{figure*}

\emph{dHybrid} is a non-relativistic code, and cannot directly test the $E^{-2}$ regime. 
However, the $p^{-4}$ dependence is common to both relativistic and non-relativistic particles; 
therefore, it is reasonable to expect that the obtained momentum spectrum may really be universal. 

By looking at the time evolution of the non-thermal ion distribution in figure \ref{fig:evo}, one notes that the spectral slope remains almost constant in time, while the high-energy cut-off, well-fitted by an exponential $\propto \exp{(-E/E_{max})^{\tau}}$, with $\tau\sim 1.5$, moves to larger and larger energies at later and later times. 
We stress that, since our simulation box is very large in the $x$-direction, $E_{max}(t)$ is not artificially limited by the finite size of the simulation until $t\simeq 2000\omega_c^{-1}$ at least, but it is only determined by the acceleration time.
We will discuss the properties of ion diffusion and the time evolution of $E_{max}$ in a more quantitative way in a forthcoming work.

\section{Supra-thermal particles}\label{sec:therm}
\begin{figure*}\centering
\includegraphics[trim=0px 0px 0px 0px, clip=true, width=\textwidth]{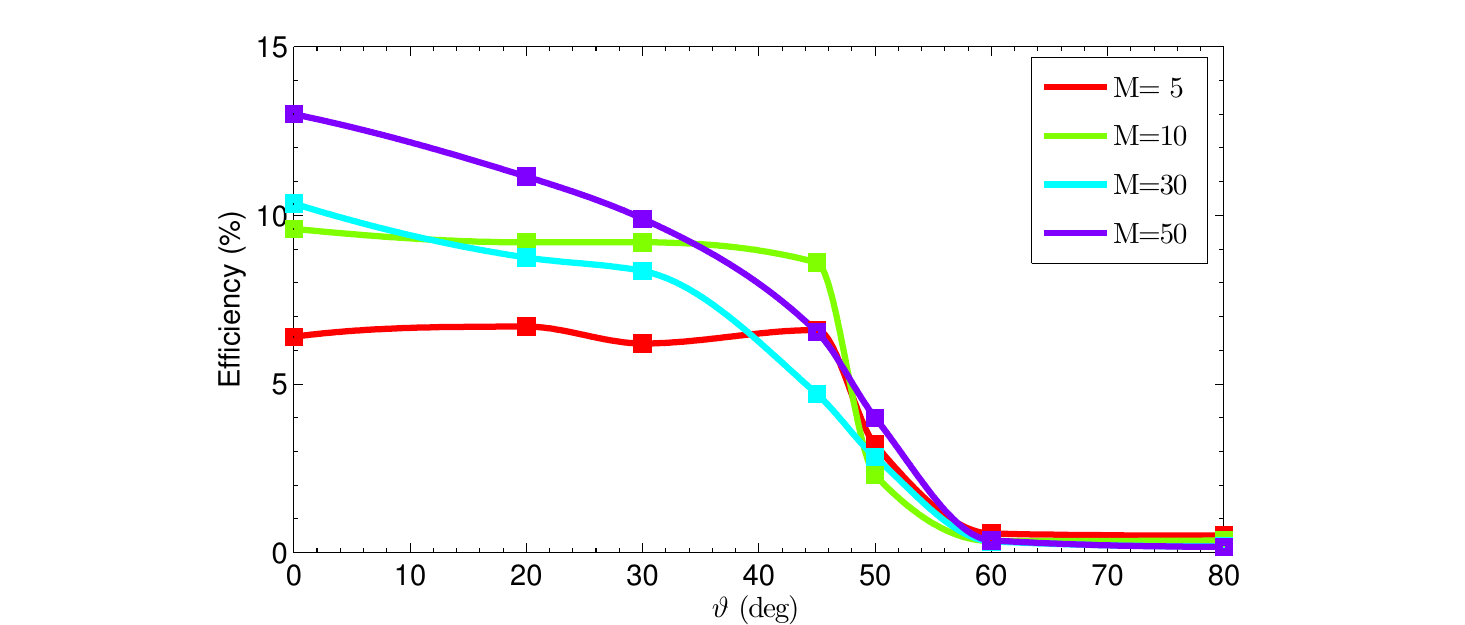}
\caption{\label{fig:eff}
Acceleration efficiency, defined as the fraction of the post-shock energy density in particles with $E\geq 10 E_{sh}$, at $t=200\omega_c^{-1}$, for several shock inclinations and Mach numbers.
There is a very significant drop in the acceleration efficiency for $\vartheta\gtrsim 45^{\circ}$, and the largest efficiency is achieved for fast, parallel shocks.
\emph{A color version is available in the online journal}.
 }
\end{figure*}

While far downstream the transition between the Maxwellian and the power-law tail is very sharp, immediately behind the shock the situation is quite different.
As shown in figure \ref{fig:therm} (red curve), the post-shock ion distribution shows a ``bridge'' of about one order of magnitude in energy (from a few $E_{sh}$ up to $\sim 10 E_{sh}$), which can be fitted with a quite steep power-law $\propto E^{-3}$. 
This spectral feature at \emph{supra-thermal} energies gradually disappears when moving downstream (figure \ref{fig:therm}).
Far downstream ($\gtrsim 3000c/\omega_p$ behind the shock) the spectrum can be clearly separated into the thermal and the non-thermal distributions.
  
The supra-thermal region is important because it contains information about the thermalization of the incoming plasma stream.
It also provides a hint that behind the shock there is a pool of particles with mildly non-thermal energies, which is crucial for understanding the problem of \emph{particle injection}, namely the determination of the conditions required for some particles to take part in the DSA process.

In the past decades, many efforts have been dedicated to the study of the effects of efficient acceleration at shocks (see references in section \ref{sec:intro}), and in particular to the investigation of the dynamical back-reaction of CRs on the plasma flow and on the electromagnetic field.
Excellent reviews on these \emph{CR-modified shocks} are, e.g., \cite{drury83,jones-ellison91,malkov-drury01}. 

Any non-linear model of DSA, as well as any phenomenological model that aims to explain the non-thermal emission from astrophysical shocks, requires the knowledge of the fraction of particles injected in the power-law tail.
Such a normalization can be worked out only within a self-consistent description of the shock transition, both in terms of electromagnetic fields and particle distribution. 
Kinetic simulations can provide this information, which may then be fed into models that deal only with time- and length-scales much larger than the background  plasma ones.
An assumption common to all of the non-linear approaches to DSA is that the shock is considered an infinitesimally thin transition where both isotropization and particle injection occur. 

A popular way of dealing with injection is represented by the \emph{thermal leakage} model, which assumes that particles living sufficiently far in the tail of the downstream Maxwellian, and sufficiently close to the shock, have gyroradii large enough to recross the shock in one orbit \citep[see, e.g.,][]{eje81,malkov97,KJG02,bgv05}. 
Conversely, some authors interpreted the output of their kinetic simulations as evidence that most of the particles are injected by being reflected at the shock surface \citep[see][and references therein]{gg13}, so that the idea of ``leaking'' may be misleading.

The asymptotic (far downstream) ion distribution found in our simulations shows that the fraction of injected particles can be effectively \emph{parametrized} by defining a threshold energy, $E_{inj}$, which marks the boundary between the thermal and non-thermal distributions (see the bottom panel of figure \ref{fig:therm}). 
The number of non-thermal particles can be estimated by a simple continuity argument, the CR spectrum being steep enough to be dominated by number by its lowest-energy boundary. 
If $f_{th,d}(E)$ is the Maxwellian distribution with the downstream temperature $T_d$, the total number of injected ions must be calculated at $\approx f_{th}(E_{inj})$. 

In the context of the thermal-leakage model, the injection momentum $p_{inj}=\sqrt{2 m E_{inj}}$ is expressed as a multiple of the downstream thermal momentum, namely 
\begin{equation}\label{eq:xi_inj}
p_{inj}=\xi_{inj}p_{th}; \quad p_{th}=\sqrt{2mk_B T_d}.
\end{equation}
For a strong shock, $p_{th}=\frac{4\gamma\sqrt{\gamma-1}}{(\gamma+1)^2}m v_{sh}\simeq 0.77 m v_{sh}$, with $\gamma=5/3$.
Note that here $v_{sh}$ is the velocity of the upstream fluid in measured in the downstream reference frame, while the jump conditions are usually calculated in the shock frame.
From figure \ref{fig:therm} we infer $E_{inj}\simeq 4- 5 E_{sh}$, and thereby
\begin{equation}
 \xi_{inj}\simeq 3- 3.5 \,. 
\end{equation}
A necessary caveat is that, even if we account for more than 2 orders of magnitude in the non-thermal tail, the energy spectrum of the CRs accelerated, e.g., in SNR shocks, is expected to extend into the relativistic regime for several decades. 
Since the ``real'' acceleration efficiency cannot be either larger than 1, or lower than what we find (15--20\%), and since for a $p^{-4}$ spectrum the CR energy per decade is almost constant, a simple energetic argument suggests the normalization of a ``real'' CR spectrum to be smaller by a factor of $\sim\log_{10}(p_{max}/p_{inj})\simeq 5-10$, depending on $v_{sh}/c$, on $p_{max}$, and on the actual slope of the CR spectrum. 

The dependence on $\xi_{inj}$ of the fraction of injected particles, $\eta$, is quite strong and reads:
\begin{equation}
\eta\simeq\frac{4\pi p_{inj}^3 f_{th}(p_{inj})}{n} \propto \xi_{inj}^3\exp{(-\xi_{inj}^2)}.
\end{equation}
An increase of $\sim 0.2- 0.5$ in the value of $\xi_{inj}$ determined in self-consistent kinetic simulations should effectively compensate for the necessarily limited extension of the obtained non-thermal spectra.
All these effects considered, we can conclude that kinetic simulations suggest that for parallel non-relativistic shocks a fraction of about $10^{-3}- 10^{-4}$ of the particles crossing the shock is injected into the DSA process, and that the injection momentum is $p_{inj}\simeq 3- 4 \,p_{th}$. 
The actual mechanisms leading to ion injection will be discussed in a forthcoming paper.

\section{Acceleration efficiency}\label{sec:eff}
A crucial question is how ion injection and acceleration depend on shock strength ($M$) and obliquity, defined by the angle $\vartheta$ between the normal to the shock and the background magnetic field $\bf{B}_0$.

To address this question, we have run several hybrid simulations with box size $(L_x,L_y)=(40000, 500) [c/\omega_p]^2$, two cells for ion skin depth and 4 particles per cell; 
the time-step is chosen as $\Delta t=(0.01/M) \omega_c^{-1}$, in order to allow for a constant Courant number ($=v_{sh}\Delta t/ \Delta x$) throughout all the runs.
The shock evolution is followed until $t=200\omega_c^{-1}$ in all the cases.
In order to capture the potential role of the filamentation instability, large computational boxes are needed \citep[][hereafter CS13]{filam}, and large Mach number shocks require quite small time steps to enforce energy conservation, which is not guaranteed in explicit hybrid methods \citep[see, e.g.,][]{Giacalone+93,gargate+07}.

We consider several obliquities, corresponding to $\vartheta=0\deg,20\deg,30\deg,45\deg,50\deg,60\deg,80\deg$ and several Mach numbers, $M=5,10,30,50$, spanning a large parameter space meant to account for the weak shocks in the Solar System and in the intra-cluster medium, but also for the strong shocks relevant for SNRs.
 
We checked the convergence of our results against the number of particles per cell, the box size, and the time and space resolution.
This work extends the investigation of GS12 to much larger boxes, crucial not to artificially limit the growth of $E_{max}$ (because of ions escaping from the upstream boundary), and to account for the effects of the filamentation instability.
Also, the simulations presented here have been run at a fixed Courant number ($\mathcal{C}\sim v_{sh}\Delta t/\Delta x$), rather than at a fixed $\Delta t$ as in GS12: 
running a survey on $M$ at a fixed time step artificially suppresses ion acceleration at strong shocks, for which energy is systematically worse conserved.
Finally, GS12 did not distinguish between supra-thermal and non-thermal ions, thereby the acceleration efficiency they quote cannot be directly interpreted as proper DSA efficiency.  
For all these reasons, we argue that figure \ref{fig:eff} provides a physically-consistent picture of CR acceleration efficiency at non-relativistic shocks, while the results of GS12 are less conclusive. 
The present analysis also extends the range of $M$ up to 50, the range of $\vartheta$ up to 80$\deg$, and supports  the results obtained in the 2D configuration with unprecedentedly-large 3D simulations.

\begin{figure*}\centering
\includegraphics[trim=0px 0px 0px 0px, clip=true, width=\textwidth]{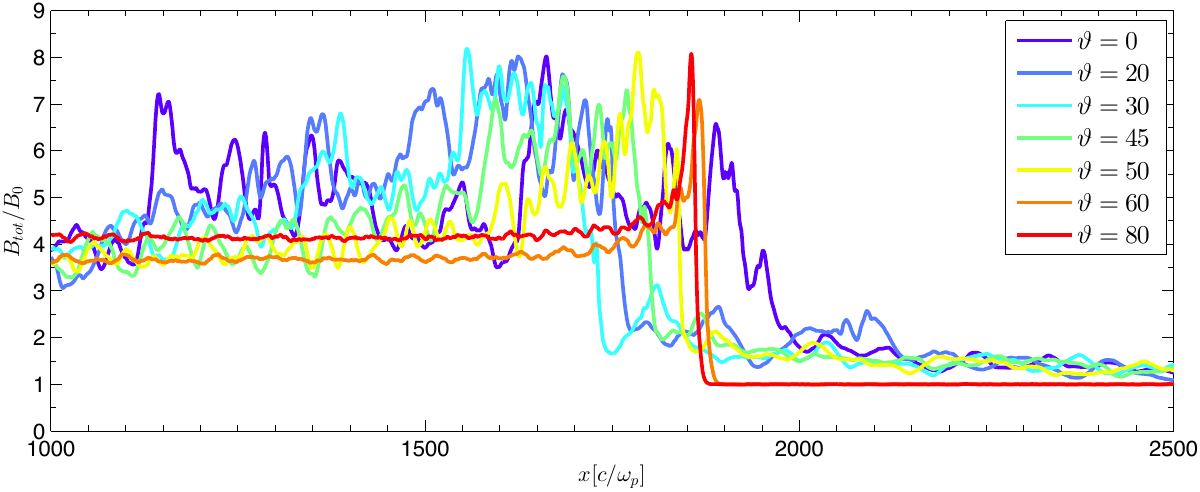}
\caption{\label{fig:Btot-theta}
Magnitude of the magnetic field for $M=30$ shocks with different values of $\vartheta$, as in the legend.
Quasi-parallel configurations show magnetic field amplification in the upstream, up to factors of $\delta B/B_0=3-4$ ahead of the shock (at $x\approx 2000 c/\omega_p$), which implies quite large post-shock magnetic fields ($B_{tot}=6-10B_0$, on average, with local peaks of $B_{tot}=10-30B_0$).
At quasi-perpendicular shocks, instead, the downstream magnetic field is just the one achieved by simple compression, apart from a narrow spike immediately behind the discontinuity. 
\emph{A color figure is available in the online journal}.
 }
\end{figure*}

In figure \ref{fig:eff} the acceleration efficiency $\ecr$ is plotted as a function of $\vartheta$, for different Mach numbers, as in the legend.
We define $\ecr$ as the fraction of the post-shock energy density in the shape of particles with energy larger than $10 E_{sh}$.
This definition, independent of the shock Mach number, is consistent with the fact that these particles are observed to diffuse ahead of the shock (see returning particles in the top panel in figure \ref{fig:therm}), and to increase their energy with time.
Alternative definitions or cuts in energy do not appreciably change  the interpretation of the results, as long as the two physical conditions above are met.

The most striking result in figure \ref{fig:eff} is that the acceleration efficiency drops above $\vartheta\sim 45\deg$. 
At quasi-parallel shocks ($\vartheta\lesssim 40\deg$), $\ecr$ is almost independent of the inclination, especially for low Mach numbers ($M=5,10$);
conversely, $\ecr$ tends to zero at high obliquities ($\vartheta\gtrsim 60\deg$), at all the Mach numbers.
Among the presented runs, the highest acceleration efficiency is achieved for $M=50$, in the parallel configuration.
For low-inclination shocks, the typical efficiency is always between 5 and 15\% at the time considered ($t=200\omega_c^{-1}$). 
Since the maximum energy is still increasing with time, these acceleration efficiencies may not have saturated, yet. 
Comparisons with longer runs show that the evolution of $\ecr$ with time is not strong, and that the value at $t=200\omega_c^{-1}$ is a good proxy of the value at later times for all the shock strengths and inclinations. 

Our simulations clearly attest to a strong dependence of the ion acceleration on the shock obliquity, favoring quasi-parallel shocks over quasi-perpendicular ones.
These results are quite different from those by GS12, where the maximum acceleration efficiency was achieved i) for $15\deg$ shocks, and ii) for $M=6$. 
The reasons for these discrepancies are: i) the shock precursor is more extended for small $\vartheta$, and is easily suppressed in simulations with small box in the longitudinal direction; ii) simulations in GS12 have fixed $\Delta t$ rather that fixed Courant numbers as those presented here; 
therefore, in GS12 the dynamics of high-energy particles is not properly accounted for at large $M$.

\section{Magnetic field amplification}\label{sec:MFA}
\begin{figure*}\centering
\includegraphics[trim=0px 80px 0px 0px, clip=true, width=.88\textwidth]{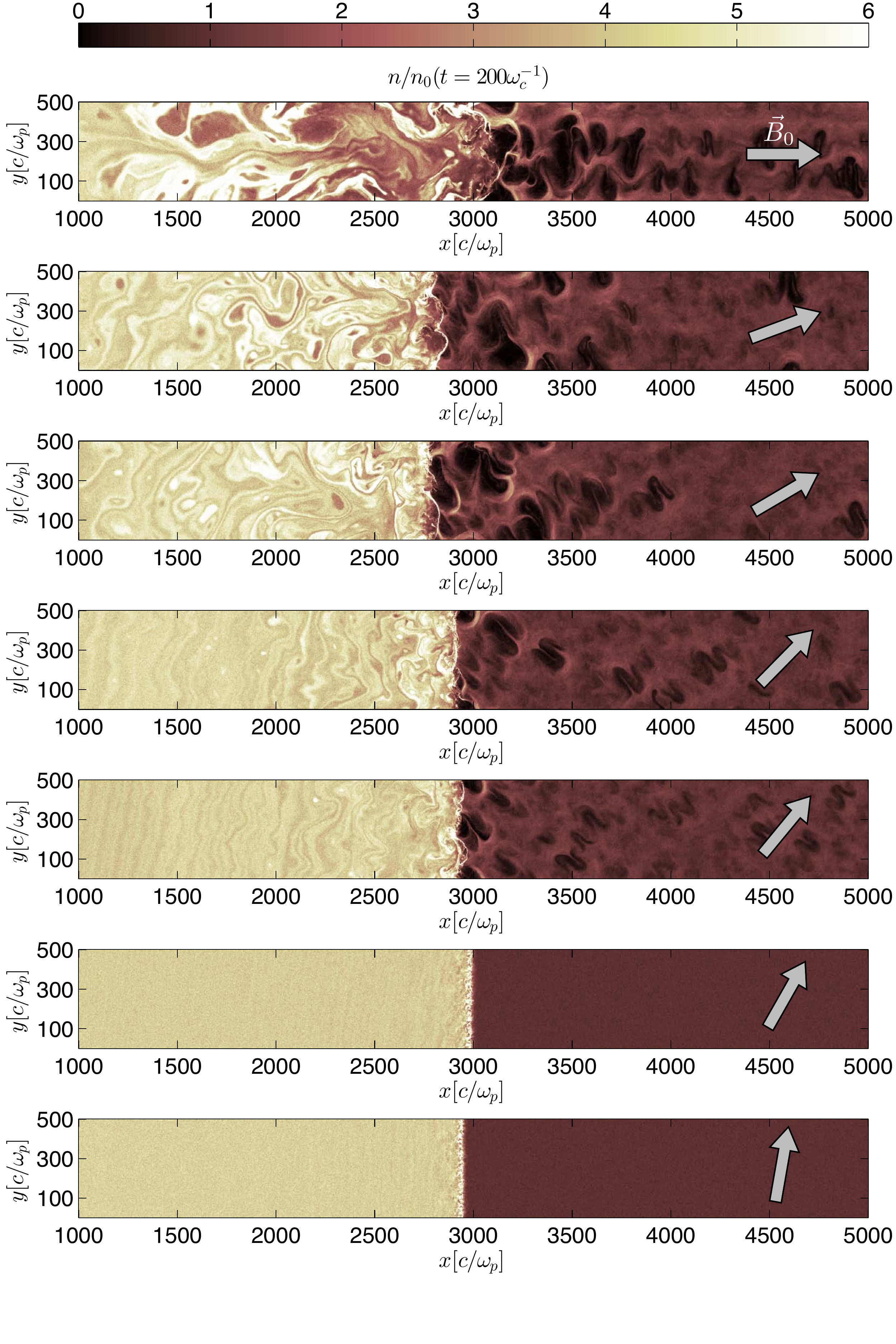}
\caption{\label{fig:rho}
Density maps for $M=50$ shocks and $\vartheta=0\deg,20\deg,30\deg,45\deg,50\deg,60\deg,80\deg$, from top to bottom, respectively. The arrow indicates the direction of the initial magnetic field.
\emph{A color figure is available in the online journal}.
 }
\end{figure*}

\begin{figure*}\centering
\includegraphics[trim=0px 80px 0px 0px, clip=true, width=.88\textwidth]{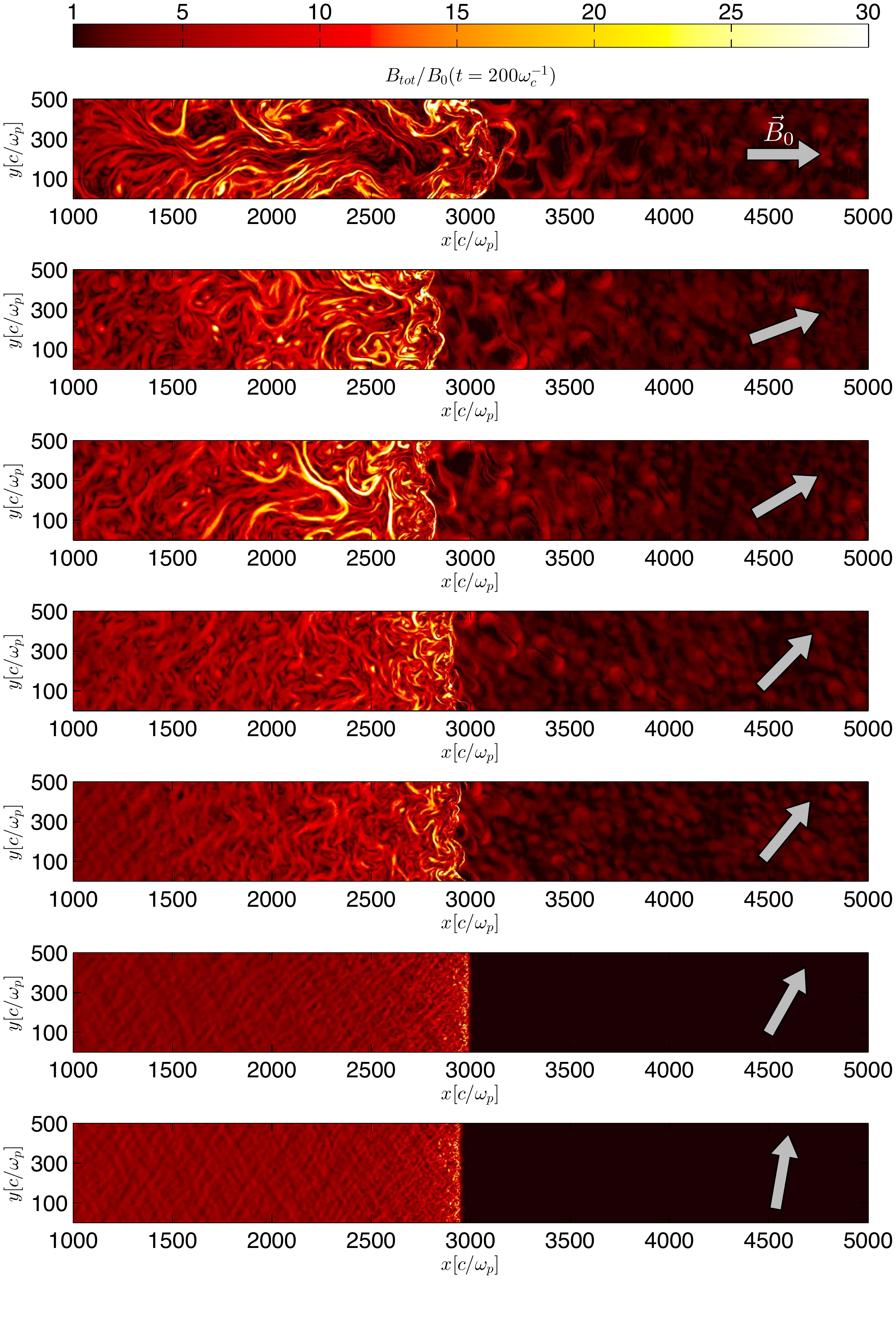}
\caption{\label{fig:Btot50}
Magnitude of the magnetic field for $M=50$ shocks and $\vartheta=0\deg,20\deg,30\deg,45\deg,50\deg,60\deg,80\deg$, from top to bottom, respectively.
\emph{A color figure is available in the online journal}.
 }
\end{figure*}

\begin{figure*}\centering
\includegraphics[trim=0px 80px 0px 0px, clip=true, width=.88\textwidth]{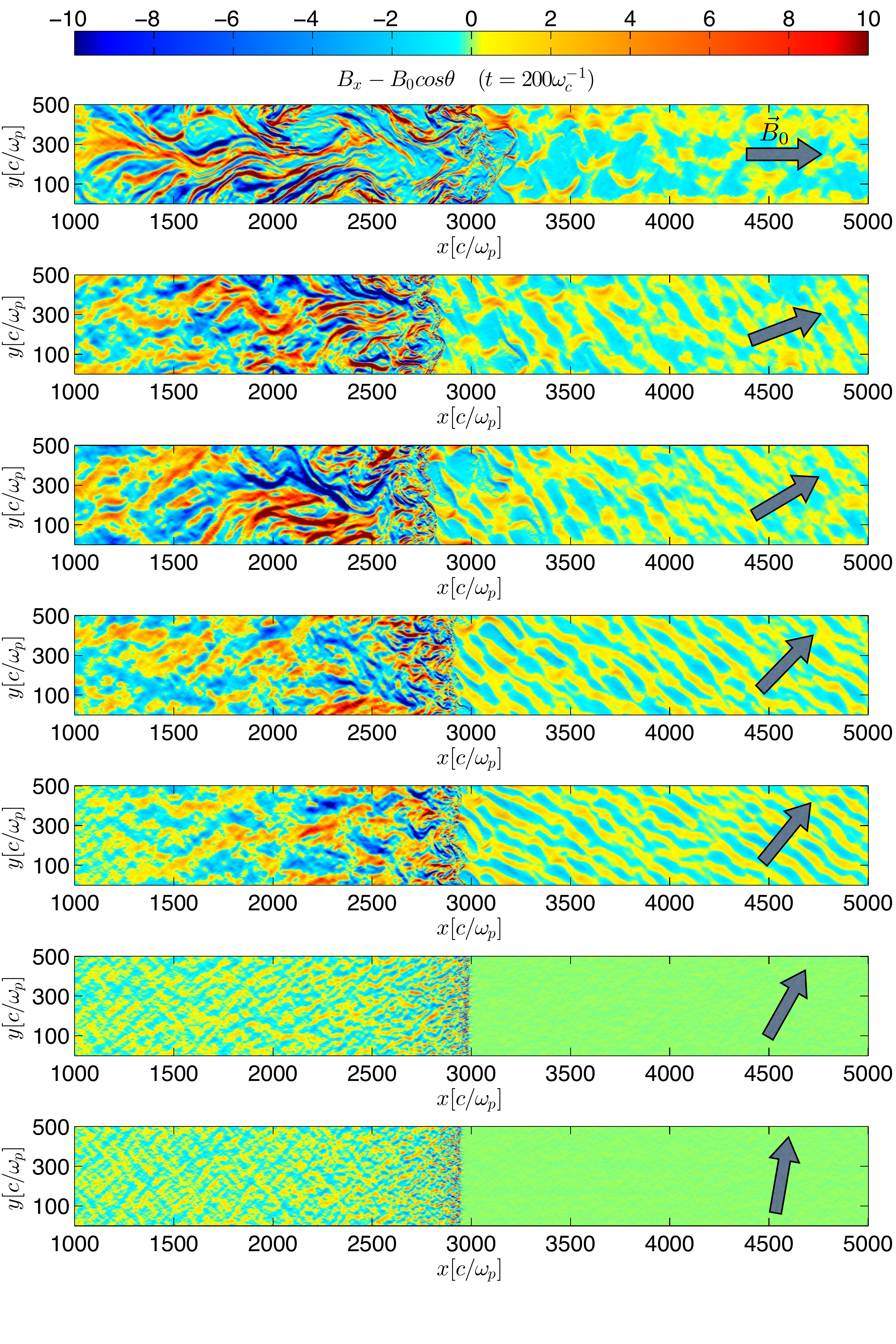}
\caption{\label{fig:Bx}
Parallel magnetic field maps for $M=50$ shocks and $\vartheta=0\deg,20\deg,30\deg,45\deg,50\deg,60\deg,80\deg$, from top to bottom, respectively. 
The color code is centered on the initial value to facilitate the comparison among different panels.  
\emph{A color figure is available in the online journal}.
 }
\end{figure*}

\begin{figure*}\centering
\includegraphics[trim=0px 80px 0px 0px, clip=true, width=.88\textwidth]{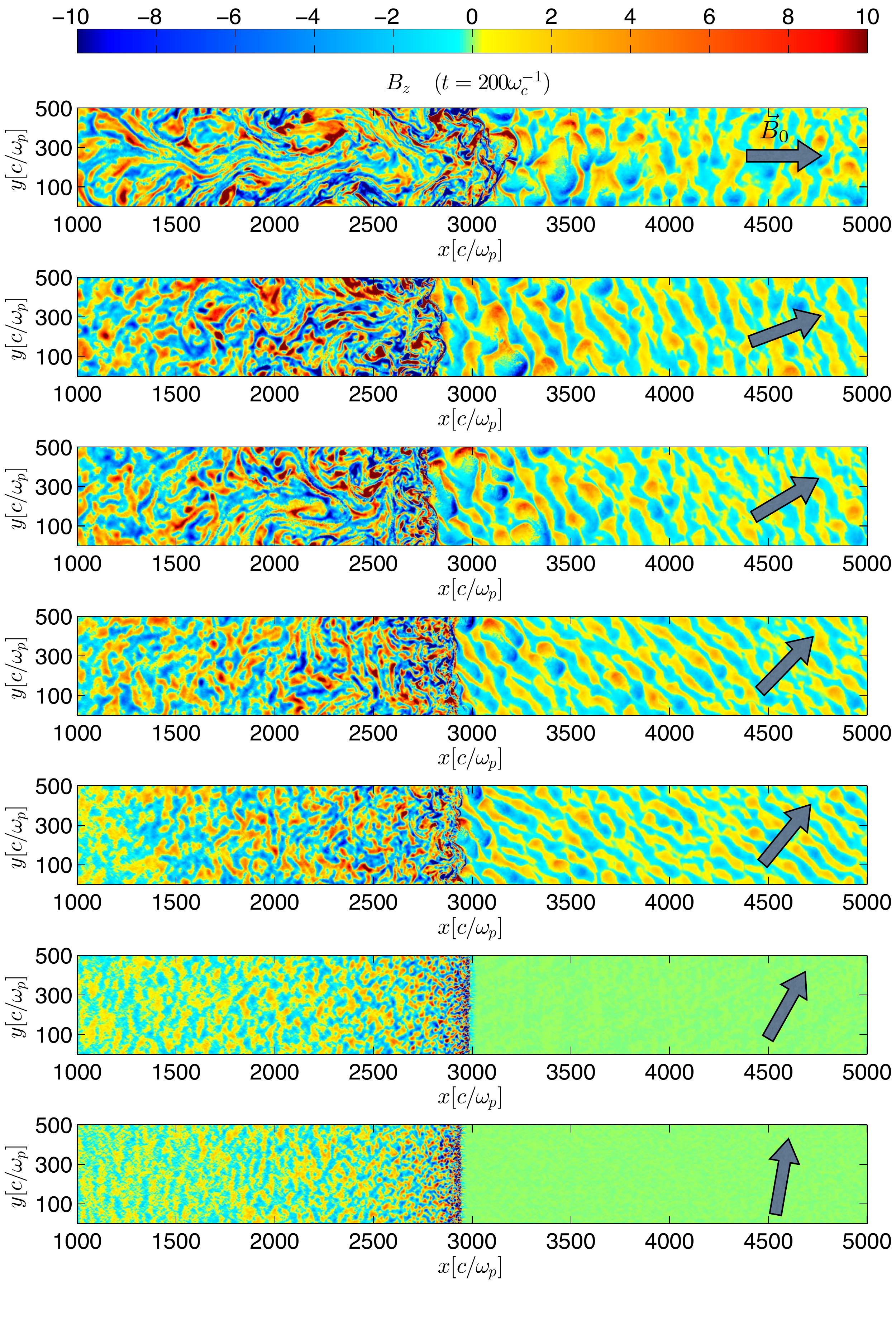}
\caption{\label{fig:Bz}
Out-of-plane magnetic field maps for $M=50$ shocks and $\vartheta=0\deg,20\deg,30\deg,45\deg,50\deg,60\deg,80\deg$, from top to bottom, respectively.
\emph{A color figure is available in the online journal}.
 }
\end{figure*}

Being driven by CR-induced instabilities, the magnetic field amplification naturally correlates with the abundance of non-thermal ions.
At quasi-perpendicular shocks, where acceleration is inefficient, the downstream magnetic field is ordered (along $\bf y$) and only $\sim 4$ times larger than the initial one, consistent with the simple compression of the transverse component of the field.
At quasi-parallel shocks, instead, the field is first amplified in the precursor, then compressed at the shock, and eventually further enhanced by turbulent motions triggered by the shock corrugation driven by the CR precursor (CS13).
These different phenomena are illustrated in figure \ref{fig:Btot-theta}, which shows the profile of the magnitude of the magnetic field, averaged over the transverse direction, for shocks with different inclinations and the same $M=30$.

The magnetic field topology is illustrated for $M=50$ and different shock inclinations in figures \ref{fig:Btot50}--\ref{fig:Bz}.
The total magnetic field map (figure \ref{fig:Btot50}) matches very well the density pattern shown in figure \ref{fig:rho}.
In the parallel and quasi-parallel configurations, the filamentation instability (in the upstream) and the Richtmeyer--Meshkov instability (at the shock transition) produce the characteristic pattern of cavities and filaments discussed by \cite{rb12} and CS13.
Filaments always develop in the direction of the initial magnetic field ${\bf B}_0$, being produced by field-aligned currents.
For high inclinations, no magnetic modes are excited to non-linear levels in the upstream, and the shock discontinuity results much sharper than in the quasi-parallel cases.
The physical explanation of this phenomenon is outlined in section \ref{sec:SDA}.

Figures \ref{fig:Bx} and \ref{fig:Bz} show the parallel and transverse (out of plane) components of the magnetic field;
the ordered parallel component $B_0\cos\vartheta$ is subtracted from $B_x$ to facilitate the comparison among different obliquities. 
As discussed by CS13, when CR acceleration is efficient, the amplified field is coiled around the upstream cavities, and turbulent in the downstream.
At quasi-perpendicular shocks, instead, the downstream  field is mainly oriented along ${\bf y}$, except in a thin layer (a few times the gyroradius of downstream thermal ions) behind the shock.

\begin{figure*}\centering
\includegraphics[trim=0px 0px 0px 0px, clip=true, width=.8\textwidth]{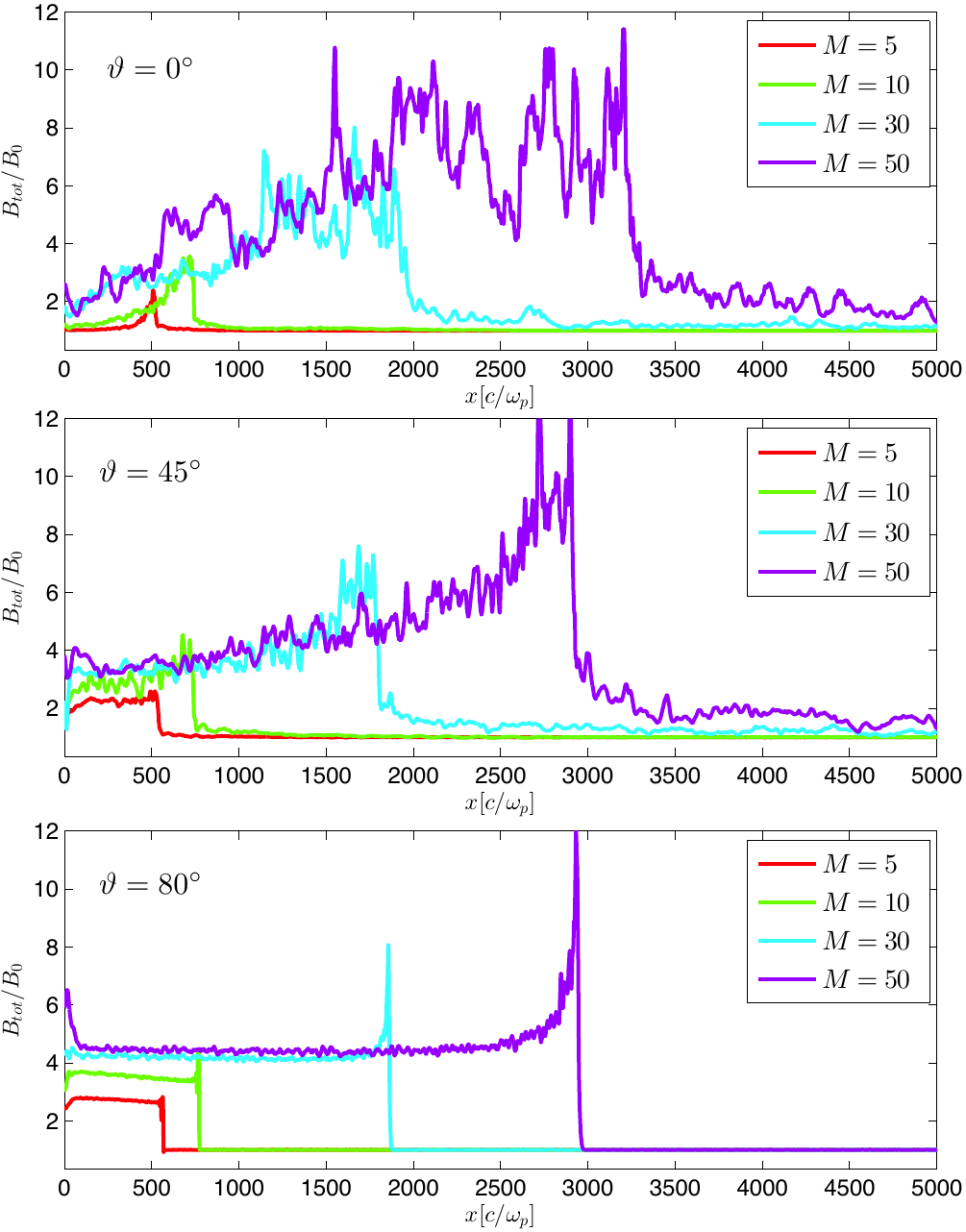}
\caption{\label{fig:Btot}
Magnitude of the magnetic field, averaged in the transverse direction, for three different inclinations ($\vartheta=0\deg,45\deg,80\deg$, from top to bottom, respectively).
Different curves correspond to the same $t=200\omega_c^{-1}$ and different Mach numbers, as in the legends.
\emph{A color figure is available in the online journal}.
 }
\end{figure*}

In the downstream, the spatial profile  is quite different for parallel and oblique cases.
Quasi-perpendicular shocks show a spike of strong field immediately behind the shock \citep[the magnetic overshoot associated with compression due to ion gyration, see, e.g.,][]{aa88}, after which the field relaxes to its asymptotic value, according to the Rankine-Hugoniot condition $B_{\perp,d}=rB_{\perp,0}$. 
In parallel shocks, instead, the region with large field ($B_d\sim 10B_0$) is much more extended, because of the amplification granted by the turbulent dynamo mechanism triggered by the corrugation of the shock surface (CS13).
Eventually, the magnetic field has to drop far downstream, in order to satisfy the asymptotic Rankine-Hugoniot condition $B_{\perp,d}=0$.
The transverse magnetic field, which accelerated ions build up by winding and amplifying the initial field, must relax on some \emph{damping} length-scale, which is found to be larger than the typical diffusion length of the accelerated ions.   

The amount of magnetic field amplification depends also on the shock strength, as illustrated by figure \ref{fig:Btot}, which shows the profile of $B_{tot}$ for shocks with different inclinations and Mach numbers.
Low Mach number shocks show the same trend outlined for the $M=50$ case, but the $\delta B/B$ factor is proportionally lower, both upstream and downstream.
The fact that magnetic field amplification becomes more and more effective for stronger shocks is encouraging for large amplification factors inferred in young SNRs, whose fast shocks are characterized by Mach numbers as large as $\sim 50-500$.

\subsection{Observational consequences: SN1006}
\begin{figure*}\centering
\includegraphics[trim=0px 340px 0px 0px, clip=true, width=.9\textwidth]{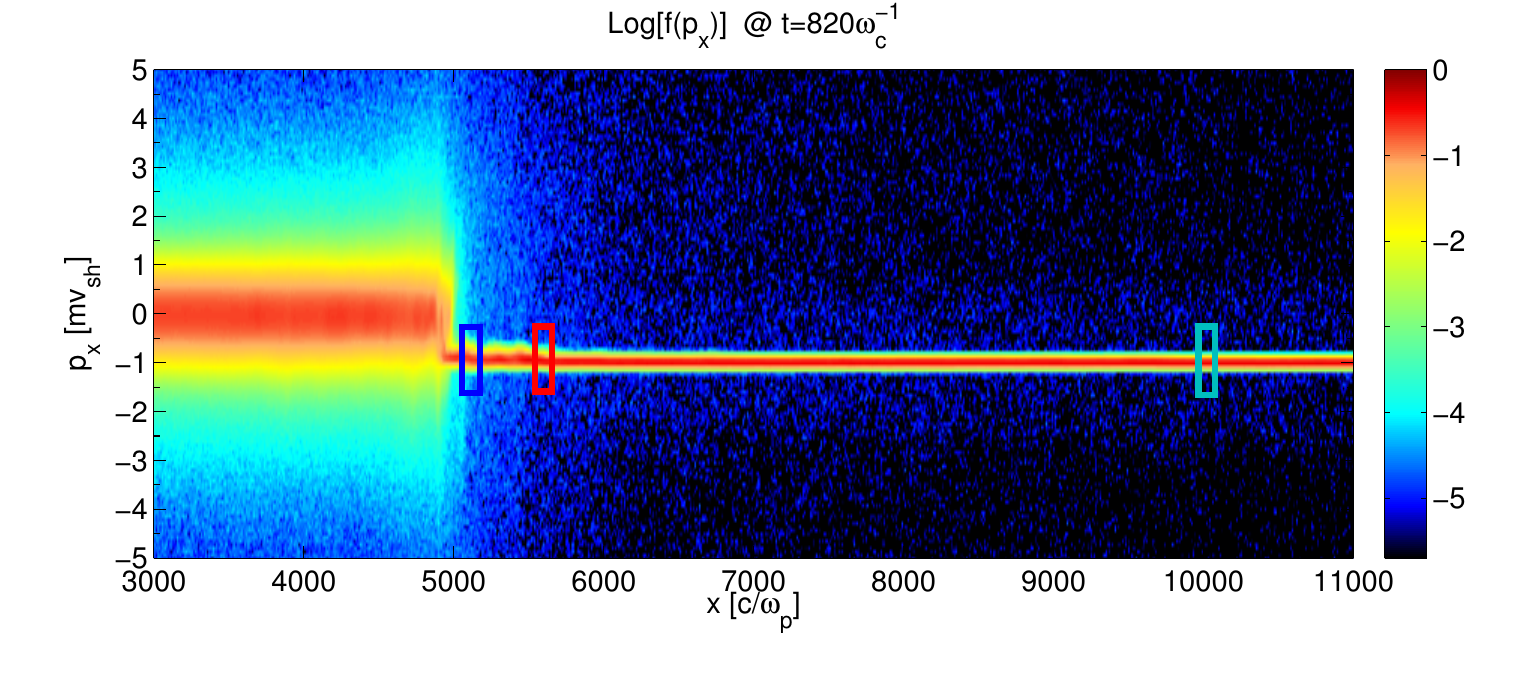}
\includegraphics[trim=0px 50px 0px 340px, clip=true, width=.9\textwidth]{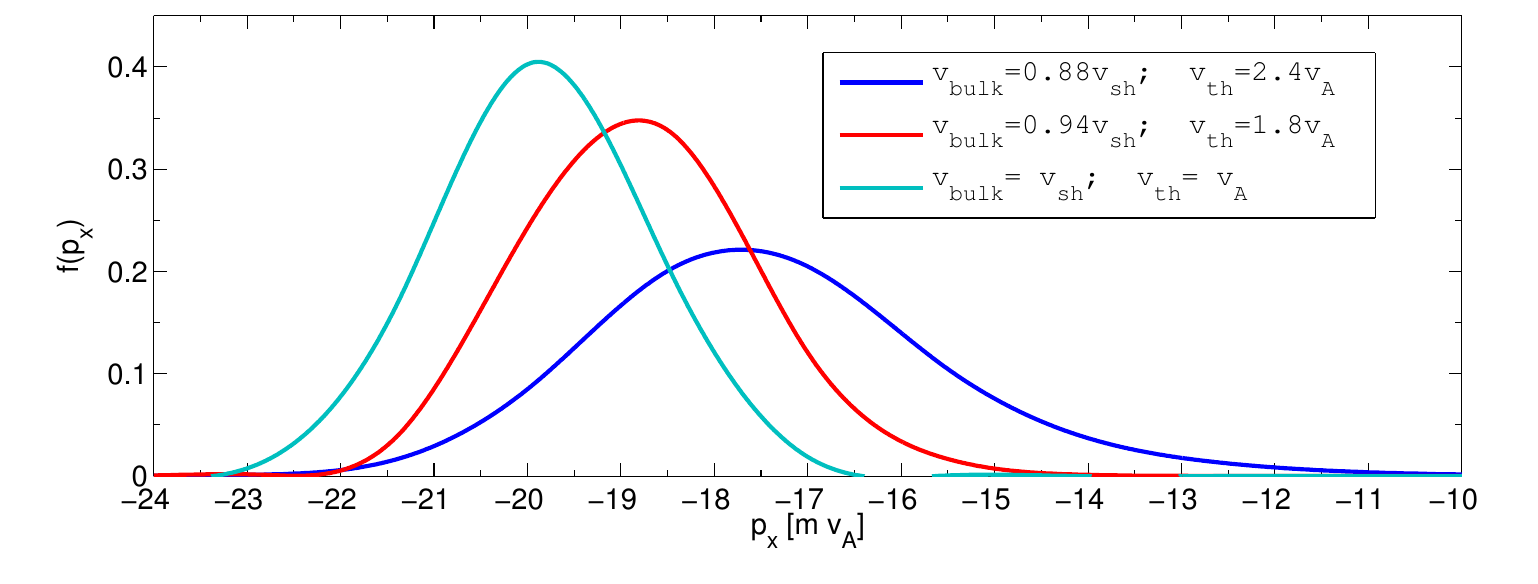}
\caption{\label{fig:prec}
\emph{Top panel}: ion $p_x$ distribution, as a function of position, for a parallel shock with $M=20$. 
The upstream beam of cold ions is converted into a hot Maxwellian distribution at the shock, around $x=5000c/\omega_p$.
\emph{Bottom panel}: ion distribution in $p_x$ for the three upstream locations in the top panel.
The initial flow has $v_{bulk}=v_{sh}=-20v_A$ and thermal spread $v_{th}=v_{a}$. 
When approaching the shock, the upstream fluid is slowed down because of the CR pressure, and heated up because of both adiabatic and turbulent heating.
\emph{A color version is available in the online journal}.
 }
\end{figure*}

\begin{figure*}\centering
\includegraphics[trim=0px 0px 0px 0px, clip=true, width=1\textwidth]{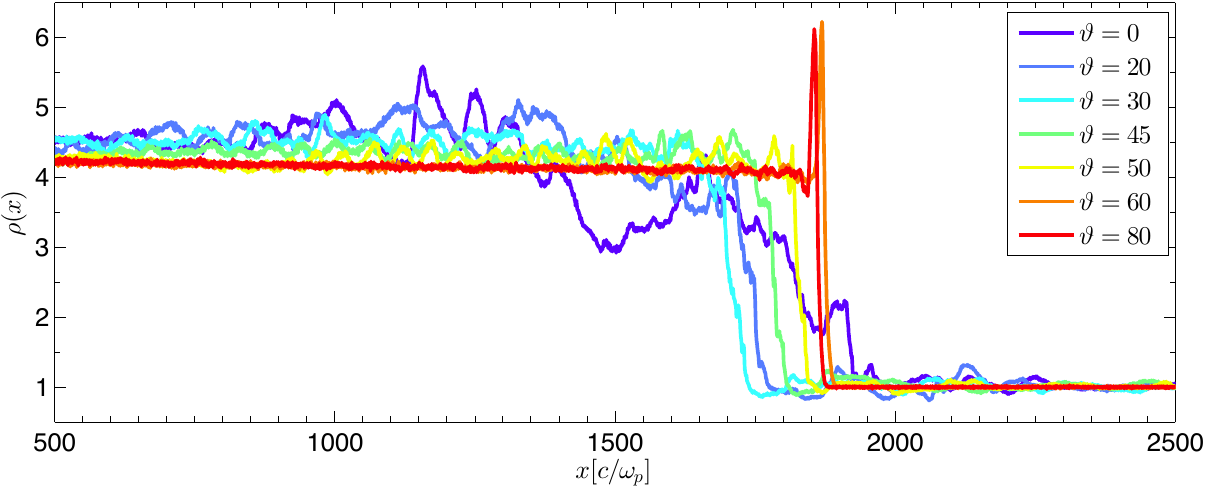}
\caption{\label{fig:rho-theta}
Density profile (integrated along the transverse direction) at $t=200\omega_c^{-1}$ for $M=30$ shocks with inclinations as in the legend.
Notice that for quasi-parallel shocks, where CR acceleration is efficient, the shock thickness is broader and the total compression ratio (i.e., the ratio between the density far downstream and far upstream) is about 4.2--4.4, systematically larger than for very oblique shocks.
\emph{A color version is available in the online journal}.
 }
\end{figure*}

The strength and the degree of isotropy of the downstream magnetic field have observational consequences in the intensity and in the degree of polarization of the synchrotron emission detected from astrophysical shocks.

The Galactic field is inferred to be coherent on scales of few tens of pc, comparable with the diameter of young SNRs.
It is thereby legitimate to wonder whether SNR blast waves might allow to test different field obliquities in different regions of the same object.

In spite of the fact that it is usually hard to determine the actual orientation of the Galactic magnetic field at SNR locations, \cite{reynoso+13} have exploited state-of-the-art radio observations to claim evidence for the bilateral shape of the non-thermal emission of SN 1006 to be consistent with a polar cap geometry.
In other words, the two lobes showing more prominent (radio and X-ray) synchrotron emission are inferred to be orthogonal to the direction of the local Galactic magnetic field.
Very interestingly, these polar caps also show a low degree of polarization in the radio band. 
Conversely, in regions with low or null non-thermal X-ray emission the degree of polarization is as high as 70\%, a value consistent with an ordered magnetic field, inferred to be perpendicular to the shock normal.
The observations by \cite{reynoso+13} agree very well with our finding large, turbulent magnetic fields in the downstream of parallel shocks, and simply-compressed fields at quasi-perpendicular shocks.

Turbulent fields are likely the result of an efficient ion acceleration, so that the azimuthal variation of the degree of radio polarization in SN 1006 seems to support our findings.
Also, the observed variation of the synchrotron emissivity is consistent with a more effective magnetic field amplification in the portions of the shell where the shock is parallel.
However, the emissivity might not depend on the magnetic file strength only, but also on how efficiently \emph{electrons} are accelerated.

Electrons are indeed observed to be accelerated at oblique shocks, for instance at interplanetary shocks produced by solar coronal mass ejections \citep{WilsonIII+12}.
This empirical evidence is supported by PIC simulations \citep[e.g.,][]{ah10,rs11,Park+12}, and may rely on the fact that whistler waves, which have been demonstrated to be important for electron acceleration \citep{rs11}, are preferentially excited at low-Mach-number oblique shocks.
Whether quasi-parallel shocks can accelerate electrons is currently under investigation \citep[also see][for some recent results about the Saturn's bow shock]{Masters+13}.

As it is typically the case, the only smoking gun attesting to efficient ion acceleration at parallel shocks would be represented by $\gamma$-rays from the decay of neutral pions produced in nuclear collision between CRs and the background gas.  
Interestingly, TeV $\gamma$-rays have been detected from SN 1006 exactly in correspondence of the polar caps \citep{SN1006HESS}.
This would be a strong hint of efficient ion acceleration, if such an emission were unequivocally proved to be hadronic.
Detailed multi-wavelength studies are needed to support or disprove this scenario.

\section{Cosmic-ray--modified shocks}\label{sec:CRmod}
When acceleration is efficient, and pressure and energy density in accelerated particles become comparable with the respective bulk quantities, one enters the regime of CR-modified shocks. 
In this section we discuss the two main dynamical consequences of efficient ion acceleration, namely the formation of a shock precursor and the modification of the shock jump conditions that we observe in our simulations.

\subsection{Upstream precursor}\label{sec:precursor}
The most prominent signature of a CR-modified shock is the formation of a \emph{precursor}, in which the upstream fluid is slowed down because of the pressure in CRs diffusing back and forth across the shocks.
Indicating with $\tilde{u}$ the fluid velocity in the shock reference frame, in the stationary limit one has from mass flux conservation:
\begin{equation}\label{cons:mass}
n(x) \tilde{u}(x) = {\rm const}, 
\end{equation}
so that deceleration also leads to compression, and hence to adiabatic heating.
However, if some amplified magnetic modes were dissipated, for instance because of non-linear Landau damping, there should be an additional source of heating, usually referred to as  \emph{turbulent} (or \emph{Alfv\'en}) \emph{heating} \citep[see, e.g.,][]{mckenzie-volk82,be99,ab06}.

Let us consider a 2D simulation whose box size measures $(L_x,L_y)=(4\times 10^4, 10^3) [c/\omega_p]^2$, with two cells per ion skin depth and 4 macroparticles per cell. 
The large transverse size is chosen in order to fully account for the effects of the filamentation instability, which can develop only if the box is larger than the gyroradius of the most energetic ions, at any time.
The formation of rarefied cavities and dense filaments associated with the streaming of accelerated ions may significantly alter the results one would obtain with 1D simulations (CS13).

In figure \ref{fig:prec} we show the ion $p_x$ distribution as a function of position for $M=20$ parallel shock. 
The flow is initialized at the upstream boundary with bulk velocity $-20 v_A$ and thermal speed $v_{th}=v_A$; at $x\simeq 10^4c/\omega_p$ the ion distribution retains basically the same properties.
By looking at the bottom panel of figure \ref{fig:prec}, we see that the bulk flow becomes slower closer to the shock, and its thermal spread increases.
In particular, immediately ahead of the shock the bulk velocity is reduced by about 12\%, and the plasma temperature $T\propto v_{th}^2$ is more than 5 times larger than the initial one.

As expected from momentum flux conservation, the amount of fluid braking matches very well the energy channeled into non-thermal ions, which at the considered time is about 13\%.
The adiabatic compression in the precursor, which in filaments reaches $\delta n/n\sim 2$, is not large enough to account for such a heating:
a variation of a factor of $\sim 2$ in density would correspond to an increase of $\sim 60$\% only in $T\propto P/n\propto n^{2/3}$.
Very interestingly, the increase in the plasma pressure is comparable with the increase in the magnetic pressure, $P_B=B^2/8\pi$, which is due to the CR-induced instability.
For the same run, the magnitude of the magnetic field, averaged over the transverse direction, is $B_{tot}\approx 2-3 B_0$ immediately ahead of the shock;
the local field can be even larger (up to $\sim 6B_0$) due to filamentary inhomogeneities.

The observed rough equipartition between thermal and magnetic pressures throughout the precursor strongly suggests that the free energy in the CR streaming is converted in both channels, likely due to the turbulent nature of the perturbation. 
The net result is that the pre-shock Mach number is reduced by about a factor of 3, as a consequence of both the decrease of the bulk flow speed (by about 15\%), and the increase of $v_{th}$  (which is about 2.4 times larger than far upstream, see figure \ref{fig:prec}).
In the presence of efficient particle acceleration, magnetic field amplification and turbulent heating of the background plasma lead to weaker subshocks, and in turn to a less efficient heating of the downstream plasma (see figure \ref{fig:evo}). 

The ion temperature in collisionless shocks may be directly probed by measuring the optical H-$\alpha$ emission produced in SNRs expanding into partially-neutral media \citep[see, e.g.,][]{GLR07,neutri1}.
In particular, \cite{neutri3} showed how to extract consistent information about the length-scale and the temperature of the upstream precursor by measuring the width of the narrow Balmer line produced by charge-exchange between neutrals and ions, even in the presence of efficient CR acceleration.

\subsection{Modified jump conditions}
Another important effect on the shock dynamics is the modification of the purely-gaseous Rankine-Hugoniot jump conditions in the presence of non-thermal particles. 
Such an effect can be outlined within a simple two-fluid approach: ions are separated into a thermal population, which has an adiabatic index $\gamma$ and feels the usual entropy variation at the discontinuity, and a CR population, constituted by particles with gyroradii larger than the shock thickness, the distribution of which is approximately constant throughout the discontinuity.

In two-fluid models \citep[see, e.g.,][]{drury83,malkov-volk96}, the CR component is usually assumed to have the adiabatic index of a relativistic gas, $\gamma=4/3$.
When acceleration is efficient, the total compression ratio ($r_{tot}$, measured between upstream and downstream infinity) may become larger than 4, since it must be calculated with an effective adiabatic index $4/3\leq \gamma\leq 5/3$ (see eq.~\ref{eq:Mtilde}).
Moreover, the possible escape of accelerated particles toward upstream infinity may make the shock behave as partially radiative, thereby contributing to increase the compressibility of the downstream plasma and, in turn, leading to $r_{tot}\gg 4$.

Nevertheless, non-thermal particles alter the jump conditions in a similar way \emph{even in the fully non-relativistic case}: the very presence of energetic particles that do not obey the gaseous density (and entropy) jump at the shock is sufficient to lead to a total compression ratio larger than 4.
The relativistic equation of state for the CRs adopted in two-fluid approaches is actually nedeed to close the system of equations; conversely, in kinetic approaches to CR-modified shocks, closure is provided by the solution of the CR transport, through different techniques \citep[see, e.g.,][]{comparison}.

Very generally, the modification in the shock velocity profile can be worked out as a function of the CR acceleration efficiency by using mass and momentum conservation only. 
We have already shown how the presence of a CR-induced shock precursor leads to a weaker subshock, the compression ratio of which reads
\begin{equation}\label{eq:rsub}
r_{sub}=\frac{(\gamma+1)\tilde{M}_1^2}{(\gamma-1)\tilde{M}_1^2+2}.
\end{equation}
Here $\tilde{M_1}$ is the sonic Mach number immediately upstream of the shock, with the fluid speed calculated in the shock reference frame. 
$\tilde{M}$ is related to $M_s$ through the implicit relation in eq.\ref{eq:Mtilde}.
The stationary momentum conservation including gas and CR pressure reads
\begin{equation}\label{eq:momentum}
\rho_0\tilde{u}_0^2+P_{g,0}+P_{cr,0}=\rho_1\tilde{u}_1^2+P_{g,1}+P_{cr,1},
\end{equation}
where 0 and 1 correspond to quantities measured at upstream infinty and immediately in front of the subshock, and the subscripts $g$ and ${cr}$ refer to thermal gas and CRs, respectively.
For simplicity, we have  neglected the magnetic field pressure in eqs.~\ref{eq:rsub} and \ref{eq:momentum}, but it is straigthforward to include such a contribution in the calculation of the actual jump conditions \citep[see][]{jumpkin}.

We assume $P_{cr,0}=0$, and normalize all the quantities to the ram pressure $\rho_0\tilde{u}_0^2$, also introducing $\Xi=P_{cr,1}/\rho_0\tilde{u}_0^2$, so that  eq.~\ref{eq:momentum} can be rewritten as
\begin{equation}
1+\frac{1}{\gamma\tilde{M}_0^2}=\frac{r_{sub}}{r_{tot}}\lbrack 1+\frac{1}{\gamma\tilde{M}_1^2}\rbrack+\Xi\,.
\end{equation}
For strong shocks $\tilde{M}_0^2\gg 1$, hence the total compression ratio simply reads:
 \begin{equation}\label{eq:rtot}		
r_{tot}\simeq r_{sub}\frac{1+1/(\gamma\tilde{M}_1^2)}{1-\Xi} 
 \end{equation} 
For the shock shown in figure \ref{fig:prec}, one infers $\Xi\approx 0.12$ and $M_{s,1}\approx 5.7$, which corresponds to $\tilde{M}_1\approx 5.6$ (eq.\ref{eq:Mtilde}).
Plugging these values in eqs.~\ref{eq:rsub} and \ref{eq:rtot} returns $r_{sub}\approx 3.65$ and finally $r_{tot}\approx 4.23$, in good agreement with the output of the simulation.

Figure \ref{fig:rho-theta} shows the density profiles for $M=30$ shocks with different inclinations: total compression ratios are typically around 4.2--4.4 for quasi-parallel shocks, where the acceleration efficiency is about 10\% or larger, while they are systematically lower for inefficient, quasi-perpendicular shocks.
The determination of the actual value of $r_{sub}$ is more complicated because of the shock broadening induced by the filamentation instability; in any case, the stationary calculation outlined above, while providing an estimate of the asymptotic shock dynamics, is not adequate to describe the subshock structure, which is intrinsically time-dependent in the simulation reference frame (see also  in figure \ref{fig:rho-theta} the density spike present at quasi-perpendicular shocks).

CR-induced precursors and modified jump-conditions may also produce spectral features, since ions with different energies may in principle probe different compression ratios in their diffusive motion.
However, in the presented simulations, it is difficult to quantify this effect for low-energy particles, which are nominally sensitive to $r_{sub}$, because in the supra-thermal region the spectrum is steeper for other reasons (see section \ref{sec:DSA}).
On the other hand, high-energy particles, which should probe the total compression ratio $r_{tot}\approx 4.3$, are expected to have a spectrum $\propto E^{-1.43}$, hardly distinguishable from the linear prediction of $E^{-1.5}$, especially considering the effects of the exponential cut-off close to $E_{max}$.

The possibility of producing spectra significantly different from the standard DSA prediction has intrigued scientists for many years.
However, concave spectra, flatter than $p^{-4}$ at the highest energies, have never been convincingly observed in SNRs, even when CR acceleration is inferred to be efficient \citep{efficiency}.
An extension of kinetic simulations to the relativistic regime would be of primary importance to shed light on the very reason of this apparent discrepancy.

\section{DSA vs SDA}\label{sec:SDA}
\begin{figure*}\centering
\includegraphics[trim=0px 20px 0px 20px, clip=true, width=.75\textwidth]{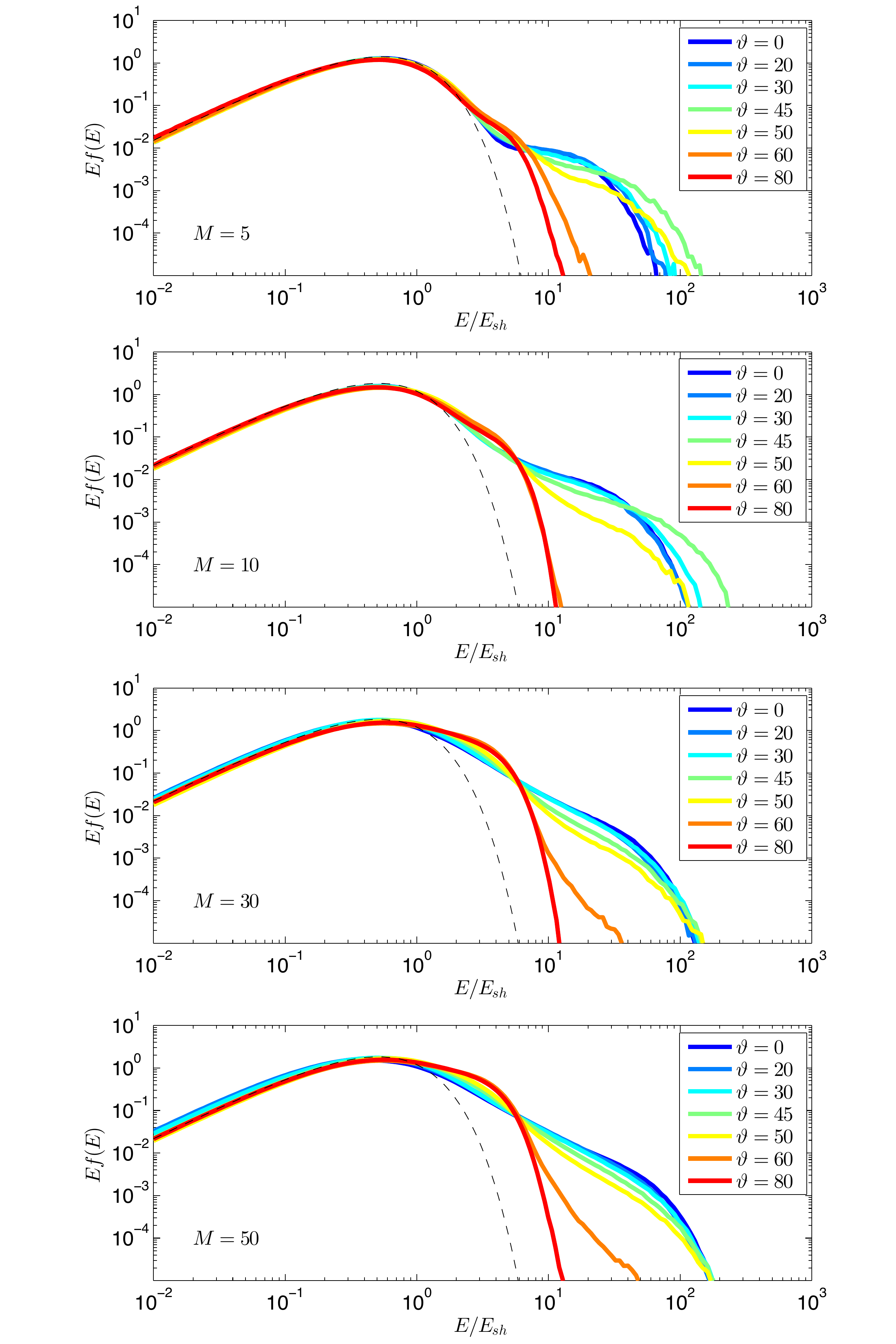}
\caption{\label{fig:M}
Post-shock particle spectra at $t=200\omega_c^{-1}$, for different Mach numbers (from top to bottom $M=5,10,30,50$), and for different shock obliquities, as in the legends.
The black dashed line represents the downstream Maxwellian.
Note how the non-thermal power-law tail develops only at low-inclination shocks.
Also note how in high-$M$, quasi-perpendicular shocks a supra-thermal ($1\lesssim E/E_{sh}\lesssim 10$) region due to SDA is present.
\emph{A color figure is available in the online journal}.
 }
\end{figure*}

The ability of a shock to accelerate ions in the non-thermal regime dramatically depends on the magnetic field topology, in the following sense. 
At quasi-parallel shocks we invariably observe a stream of ions with energies larger than a few times $E_{sh}$ propagating from the shock into the upstream plasma.
The interaction between this stream and the incoming flow excites magnetic perturbations, which favor the diffusion of ions with larger and larger energies back and forth across the shock.
This diffusion allows ions to undergo first-order Fermi acceleration, and eventually produces the universal DSA power-law distribution.

At quasi-perpendicular shocks, instead, particles do not seed the upstream plasma with self-generated waves beyond one ion gyroradius from the shock, and DSA cannot operate \emph{spontaneously}. 
We cannot exclude that, in the presence of a suitable pre-existing magnetic turbulence, DSA might operate at oblique shocks as well \citep[see, e.g.,][]{GJK94};
nevertheless, in this case the process is not guaranteed to be self-sustaining, and the achievable $E_{max}$ may depend on the largest wavelength already present in the turbulence, rather than being a dynamically-evolving quantity.  

At quasi-perpendicular shocks, ions penetrate into the upstream only while gyrating around the ordered magnetic field. 
The ions that can probe, during one gyration, the velocity jump between upstream and downstream, are accelerated via \emph{shock drift acceleration} (SDA), which allows some particles to gain energy quite rapidly.
The energy gain per cycle  is proportional to $\approx v/v_{sh}$, which means that supra-thermal particles can almost double their velocity during the first shock crossing \citep{ca75}.
However, after a few gyrations, particles experiencing SDA are advected downstream, and none of them can achieve energies larger than a few times $E_{sh}$.

\begin{figure*}\centering
\includegraphics[trim=0px 0px 0px 0px, clip=true, width=.75\textwidth]{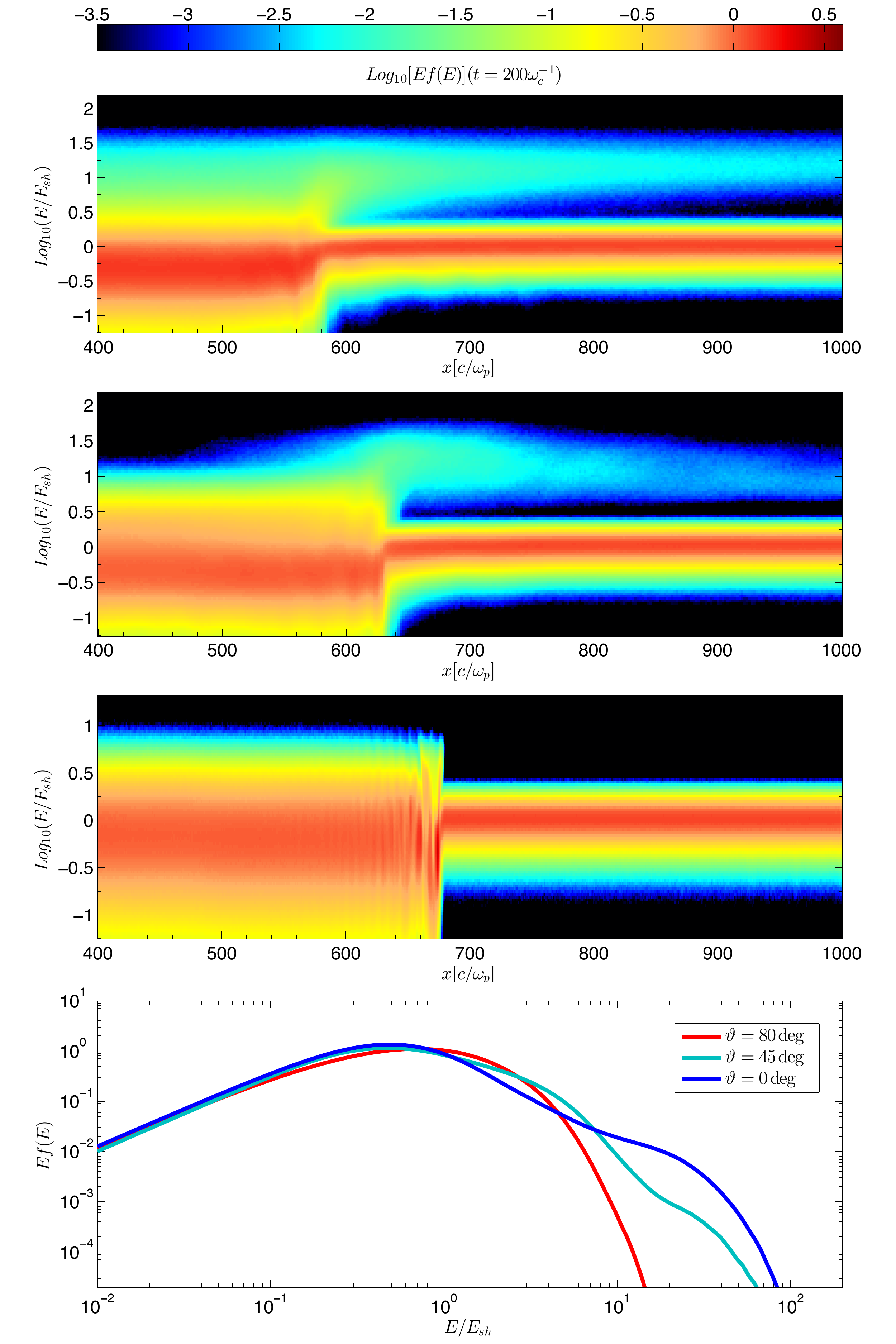}
\caption{\label{fig:3d}
Post-shock particle spectra at $t=200\omega_c^{-1}$, for 3D simulations of $M=6$ shock, for different shock obliquities.
The top three panels correspond to $\vartheta=0\deg,45\deg,80\deg$, respectively.
\emph{Bottom panel}: integrated downstream spectrum for the three cases above, as in the legend.
The non-thermal power-law tail develops only at low-inclination shocks, while at quasi-perpendicular shocks ions are only heated up by a factor of a few in energy because of SDA.
\emph{A color figure is available in the online journal}.}
\end{figure*}

\begin{figure*}\centering
\includegraphics[trim=0px 0px 0px 0px, clip=true, width=.8\textwidth]{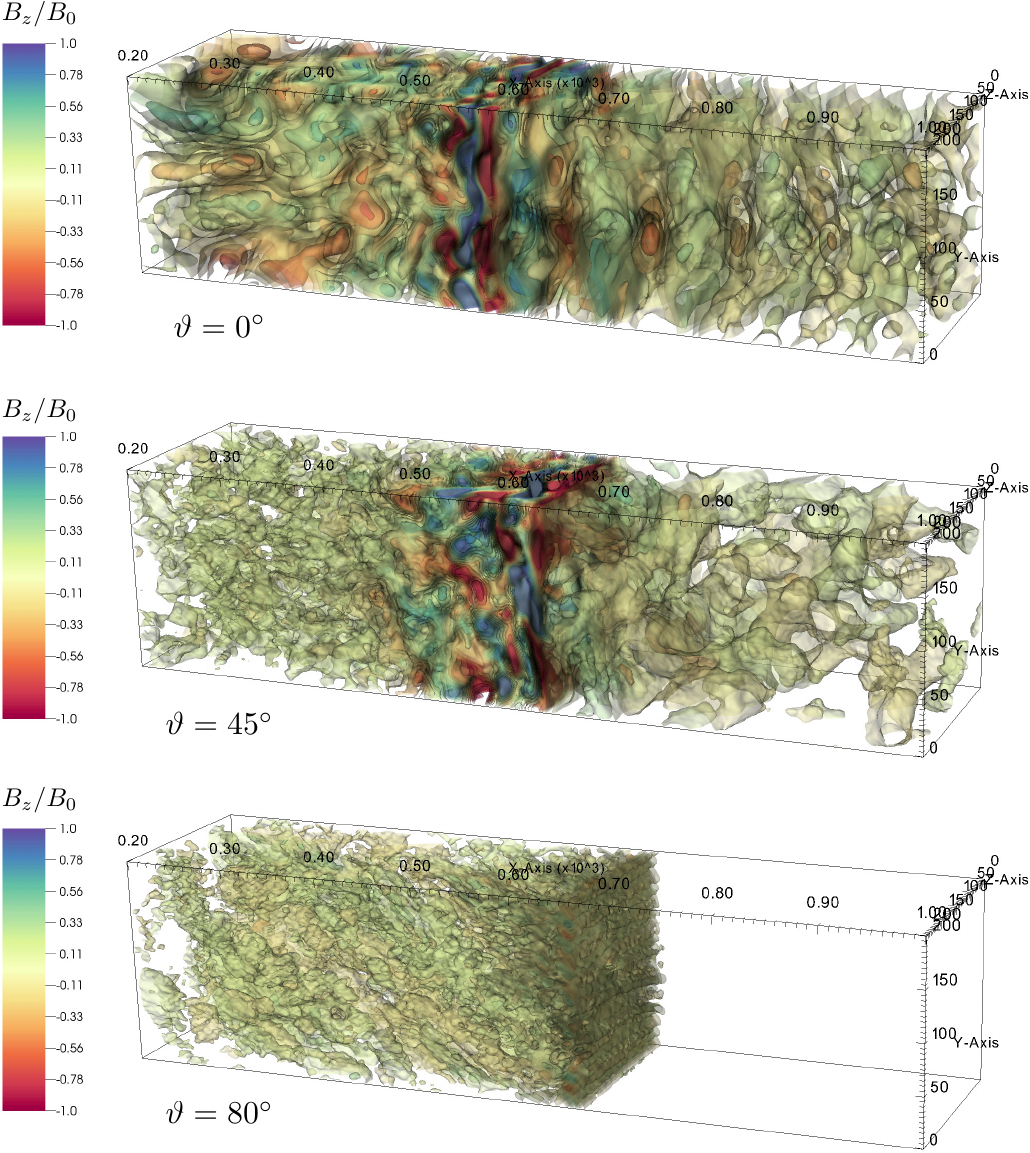}
\caption{\label{fig:3dturb}
Self-generated component of the magnetic field, $B_z$, in units of the initial field $B_0$, which lies in the $xy$-plane; the three panels correspond to $t=200\omega_c^{-1}$ for different 3D simulations (section \ref{sec:3d}) with inclinations $\vartheta=0\deg,45\deg,80\deg$ (top to bottom).
The iso-volume rendering shows 10 levels of $-1\leq B_z\leq 1$, with the respective color code in the legends.
The shock position is marked by a plane of enhanced magnetic field, around $x=600c/\omega_p$.
The amount of magnetic field amplification is very different in the parallel case, where in the upstream there are several regions with $B_z\approx B_0$, and the quasi-perpendicular case, where in the upstream $B_z\lesssim 0.1 B_0$.
Also, the magnetic field exhibits large-scale turbulent structures (both upstream and downstream) for $\vartheta=0\deg$, while it is mainly along $B_y$ for $\vartheta=80\deg$.
The $\vartheta=45\deg$ case shows intermediate properties.
\emph{A color figure is available in the online journal}.}
\end{figure*}

Also, parallel shocks develop transverse magnetic fields in the upstream, but the obliquity is never expected to be larger than $\sim 45\deg$, since in the non-linear ($\delta B/B_0\gtrsim 1$) regime the parallel component is found to grow as well, in turn always returning $B_{\perp}/B_{\parallel}\sim 1$.
However, since $B_{\perp}$ is enhanced by compression at the shock, it is hard to envision a scenario in which the downstream magnetic field does not show a prominent component transverse to the bulk fluid motion.
For this reason, some particles impinging on the shock surface are always reflected back by gyrating around this transverse magnetic field, gaining energy via SDA.
This process, strictly related to the injection problem, will be discussed in greater detail in a forthcoming paper.

In general, SDA is expected to produce a population of supra-thermal particles, which may look like a shifted Maxwellian with temperature $T_{\rm SDA}=T_d\exp\left[\frac{8r(r-1)}{(r+1)^3}\right]$, which means $T_{\rm SDA}\approx 2.2T_d$ for a compression ratio $r=4$ \citep{Park+12}.
Such a prediction is in good agreement with the ``bump'' around a few times $E_{sh}$  observed at quasi-perpendicular shocks in figure \ref{fig:M}, which shows the post-shock ion spectra for several shock strengths and inclinations.
The normalization of such a bump seems to depend on the shock strength, being more prominent for larger $M$, but this effect is only due to the fact that it takes more time to thermalize a more supersonic flow: when integrating in the downstream at a given time, the region of incomplete thermalization behind the shock (see figure \ref{fig:therm}) turns out to be more extended for high $M$.

If we are interested in CR acceleration over several orders of magnitude, though, we cannot rely on SDA only, and we have to wonder under which conditions particles can be promoted into the DSA regime.
By looking at figure \ref{fig:M}, one can see that, at fixed $M$, the total number of supra-thermal particles does not depend dramatically on $\vartheta$.
The biggest difference is that at oblique shocks reflected ions only undergo SDA, and their maximum energy is always $\lesssim 10E_{sh}$, while at quasi-parallel shocks some ions contribute to form a power-law tail, the extent of which grows with time.

We argue that the difference may be due to the geometry of the field immediately \emph{upstream} of the shock.
An ordered transverse field prevents supra-thermal ions from penetrating into the upstream for more than one gyroradius, which dramatically reduces the excitation of magnetic turbulence upstream, for two main reasons.
\begin{itemize}
\item First, the anisotropy of particles reflected at the shock is always in the direction of the flow, hence perpendicular to the magnetic field; 
in this configuration many instabilities, such as the fire-hose and the resonant streaming instability, are suppressed \citep[see, e.g.,][for a recent review]{Bykov+13}.
\item Second, and probably most important, if accelerated ions can propagate into the upstream only for a gyroradius $\sim r_0$, the time available for any magnetic perturbation to grow (i.e., the advection time $\sim r_0/v_{sh}\approx \omega_c^{-1}$) is drastically reduced with respect to the case of a parallel shock, in which energetic particles can propagate for a diffusion length $\sim v/v_{sh} r_0\gg r_0$.
\end{itemize}
Without CR-induced instabilities, the upstream magnetic field topology is left unperturbed, the shock is not corrugated, and particles cannot seed magnetic waves on larger and larger scales, that can enhance the diffusion of particles with larger and larger energies.

At parallel shocks, the upstream medium is seeded with long-wavelength perturbations, which steepen in the shock layer and lead to downstream fields that are generally oblique and that can effectively reflect incoming ions.
Nevertheless, reflected ions can fly away from the shock by spiraling around the upstream field, which may be quasi-parallel at least for a fraction of the wave period.
In other words, even if parallel shocks may \emph{often} look oblique, their time-dependent configuration, whose period is likely determined by the period of the upstream waves, is intrinsically different from the steady structure of quasi-perpendicular shocks.

Initially-parallel shocks can naturally realize both the downstream transverse field necessary for particle reflection and the upstream parallel field that allows ions to diffuse away from the shock.
Perpendicular shocks, instead, do not trigger long wavelength modes, and cannot significantly alter their initial configuration.

In summary, ion injection into the DSA process is a natural phenomenon at parallel shocks, while at oblique shocks it may require an ad-hoc pre-existent turbulence, with fluctuations of order $\delta B/B\sim 1$, possibly on scales larger than the typical gyroradius of ions with $v\sim v_{sh}$. 

\section{3D simulations}\label{sec:3d}

We also performed 3D simulations to ensure that the results outlined in the previous section are not an artifact of the reduced space dimensionality.
This is especially important because it has been suggested that cross-field diffusion, which is possible only in 3D, may play a fundamental role in ion injection \citep[e.g.,][]{BEJ95,ge00}.

We consider a shock with $M=6$, and three inclinations $\vartheta=0\deg,45\deg,80\deg$. 
The box measures $(L_x,L_y,L_z)=(2000,200,200) [c/\omega_p]^3$, with two cells for ion skin depth and 8 particles per cell; the time-step is $\Delta t=0.01 \omega_c^{-1}$.

The resulting downstream ion spectra are shown in figure \ref{fig:3d}.
The acceleration efficiency as a function of inclination follows the same trend described for the 2D cases: it is $\ecr\approx 12\%$, 3\% and 1\% for $\vartheta=0\deg,45\deg$ and $80\deg$, respectively.

Most interestingly, 3D simulations confirm the ineffectiveness of DSA at very oblique shocks, despite the exact account of the topology of the magnetic field lines. 
We do observe a large fraction of ions recrossing the shock and experiencing SDA, but no particles make it into the regime where DSA and magnetic field amplification are efficient; also, the maximum energy does not increase with time.

The 3D structure of the self-generated turbulence is shown in figure \ref{fig:3dturb} for the three inclinations above. 
More precisely, the iso-value surfaces (10 levels in the interval $B_z\in [-1,1]$) and the color code (in the legend) show the value of the component of the magnetic field orthogonal to $\bf{B}_0$ (which is initialized in the $xy$-plane). 
The shock surface, found at $x\approx 600c/\omega_p$, is always marked by an increase of $B_z$, but the upstream and the downstream structures significantly depend on the shock obliquity. 

In the parallel case, $\vartheta=0\deg$, there are several regions (also in the upstream) in which the field is amplified at levels of $\delta B/B_0\approx 1$, and the turbulence is sustained up to large distances in the downstream;
the magnetic structures have typical coherence lengths of a few tens to hundreds of $c/\omega_p$.

In the quasi-perpendicular case ($\vartheta=80\deg$), instead, $B_z$ is always less than $\sim 0.1B_0$ in the upstream, and less than $\sim 0.3B_0$ in the downstream. 
The actual magnetic field (not shown in the figure) is mainly along $\bf{y}$, and is consistently compressed in the shock transition, which is rather sharp, unlike in the parallel case.
The $\vartheta=45\deg$ case shows some intermediate features, exhibiting some magnetic structures in the upstream, even if less prominent than for the parallel configuration; 
the self-generated component of $B_z$, however, tends to fade a few hundreds of ions skin depths behind the shock.

It is worth stressing that our results are not entirely inconsistent with the previous ones that claimed ion acceleration to be efficient at perpendicular shocks, possibly thanks to cross-field diffusion.
The aforementioned results have been obtained either with Monte Carlo simulations \citep{BEJ95}, or by tracking test-particles on top of the output of hybrid simulations seeded ---by hand--- with large-scale magnetic waves \citep{ge00}.
In these cases the ion scattering is either prescribed (through the specification of a mean-free-path, in Monte Carlo simulations) or artificially enhanced, because of the presence of some pre-existing turbulence.
Our simulations, instead, allow for a self-consistent description of both the shock structure and the ion kinematics, and are long/large enough to potentially observe ions undergoing DSA \emph{in the self-generated magnetic turbulence}. 
Throughout all of the parameter space considered in this paper, for both 2D and 3D setups, we never find quasi-perpendicular shocks to be able to \emph{spontaneously} generate long-wavelength waves upstream of the shock, and, in turn, to accelerate particles into the DSA regime.

\section{Conclusions}\label{sec:conclusions}
We used hybrid simulations of collisionless shocks to calculate under which conditions ion acceleration is efficient at non-relativistic collisionless shocks.
We can summarize our main results in the following points.
\begin{itemize}
\item We showed, for the first time within a PIC/hybrid approach, that DSA can be very efficient (10--20\% in energy) at quasi-parallel shocks, producing non-thermal ion spectra \emph{with the expected universal distribution}, a power-law $\propto p^{-4}$ in momentum, i.e., $\propto E^{-1.5}$ for non-relativistic ions.
\item The extent of the non-thermal power-law tail increases with time: energetic  ions are able to self-generate magnetic turbulence to larger and larger scales, enhancing their own scattering and the probability of recrossing the shock.
The maximum energy achievable is only limited by the available computational resources (finite box size and running time).
\item The ion spectrum immediately behind the shock shows a peculiar shape at energies $2\lesssim E/E_{sh}\lesssim 10$, which deviates from both a Maxwellian and the DSA power-law.
This \emph{supra-thermal} region contains information about the post-shock thermalization, and the processes responsible for injection of particles into the acceleration mechanism.
\item At quasi-parallel shocks, the fraction of ions that undergo acceleration is $\sim 10^{-3}$ of the total. 
A good parametrization of the thermal--non-thermal transition is obtained by assuming that the DSA power-law tail is attached to the Maxwellian at the injection momentum $p_{inj}\approx 3- 4 p_{th}$, where $p_{th}$ is the downstream thermal momentum.
\item DSA occurs only at parallel and quasi-parallel shocks. 
Conversely, $\vartheta\gtrsim 45\deg$ shocks show evidence of SDA, but they are ineffective in accelerating particles above $\sim 10 E_{sh}$.
Such a limitation is intrinsic and does not depend either on the computational box, or on the reduced dimensionality of the simulations. 
3D runs turn out to be very consistent with 2D simulations in this respect.
These results apply only to initially ``laminar'' flows and ordered magnetic fields: the addition of pre-existing turbulence may modify the acceleration properties of the shocks.
\item Magnetic field amplification goes along with efficient ion acceleration: the self-generated turbulence is present only at quasi-parallel shocks, and the total amplification achievable in the simulation is larger for larger Mach numbers (up to $\delta B/B_0\approx 6$ upstream, for $M=50$).
\item The presence of accelerated ions produces an upstream precursor, in which the fluid is slowed down and compressed. 
The magnetic field amplification induced by CR-induced instabilities also leads to a non-adiabatic heating of the background plasma (turbulent heating), in such a way that plasma and magnetic field pressure are almost in equipartition throughout the precursor.
\item When acceleration is efficient, the global shock dynamics is significantly modified by the presence of non-thermal particles: in this regime of CR-modified shocks, the total compression ratio may become larger than 4, and the downstream plasma is heated up less effectively. 
\end{itemize}

This paper is conceived as the first of a series aimed to investigate several aspects of ion acceleration at non-relativistic shocks through hybrid simulations. 
Important points that are not addressed in the present work, but that will be covered in forthcoming publications, are the nature and the properties of the self-generated magnetic turbulence (also see CS13), the details of particle diffusion, and the mechanisms leading to ion injection.


\subsection*{}
We wish to thank L. Gargat\'e for providing a version of \emph{dHybrid}, P.\ Blasi, E.\ Amato and M.\ Kunz for stimulating discussions, and an anonymous referee for his/her thorough comments. 
This research was supported by NSF grant AST-0807381 and NASA grant NNX12AD01G, and facilitated by the Max-Planck/Princeton Center for Plasma Physics. 
This work was also partially supported by a grant from the Simons Foundation (grant \#267233 to AS), and by the NSF under Grant No.\ PHYS-1066293 and the hospitality of the Aspen Center for Physics.
Simulations were performed on the computational resources supported by the PICSciE-OIT TIGRESS High Performance Computing Center and Visualization Laboratory. This research also used the resources of the National Energy Research Scientific Computing Center, which is supported by the Office of Science of the U.S. Department of Energy under Contract No.\ DE-AC02-05CH11231, and Teragrid/XSEDE's Ranger and Stampede under allocation No.\ TG-AST100035.


\bibliographystyle{yahapj}
\bibliography{DSA}
\end{document}